%% file: main-file_dc-paper_submit.tex
\documentclass[final,5p,times,twocolun]{elsarticle}
\pdfoutput=1
\usepackage{lineno} 
\usepackage{amsmath}           
\usepackage{amssymb}
\usepackage{bm}                

\usepackage{graphicx}
\usepackage[space]{grffile}    
\PassOptionsToPackage{bookmarks=false}{hyperref}

\usepackage{tabularx}
\usepackage{todonotes}
\usepackage[separate-uncertainty,multi-part-units = single, allow-number-unit-breaks]{siunitx}
\usepackage{subfig}
\usepackage{url}

\modulolinenumbers[5]

\journal{NIM A}

\newcommand{\arco}{$\textrm{Ar}\textnormal{-}\textrm{CO}_2$}
\newcommand{\arcois}[1]{$\textrm{Ar}\textnormal{-}\textrm{CO}_2$ (#1)}
\newcommand{\baseline}{$\textrm{Ne}\textnormal{-}\textrm{CO}_2\textnormal{-}\textrm{N}_2$ (90-10-5)}

\newcommand{\figref}[1]{Figure~\ref{#1}}
\newcommand{\figrefbra}[1]{Fig.~\ref{#1}}
\newcommand{\Figref}[1]{Figure~\ref{#1}}
\newcommand{\secs}{Secs.~}

\newcommand{\secref}[1]{Section~\ref{#1}}
\newcommand{\secrefbra}[1]{Sec.~\ref{#1}}
\newcommand{\Secref}[1]{Section~\ref{#1}}
\newcommand{\secrefs}[1]{Sections~\ref{#1}}
\begin{document}

\begin{frontmatter}

\title{Secondary discharge studies in single and multi GEM structures}
\author[b,c]{A. Deisting\corref{corraut}}
\ead{alexander.deisting@cern.ch, Now at Royal Holloway, University of London}
\author[b]{C. Garabatos\corref{corraut}}
\ead{chilo.garabatos.cuadrado@cern.ch}
\author[e,f]{P. Gasik\corref{corraut}}
\ead{p.gasik@tum.de}
\author[b,c]{D. Baitinger}
\author[d]{A. Berdnikova}
\author[b,c]{M. B. Blidaru}
\author[c]{A. Datz}
\author[e]{F. Dufter}
\author[g]{S. Hassan}
\author[e]{T. Klemenz}
\author[e]{L. Lautner}
\author[b,c]{S. Masciocchi}
\author[e,f]{A. Mathis}
\author[i]{R. A. Negrao De Oliveira}
\author[h]{A. Szabo}
\cortext[corraut]{Corresponding author}
\address[b]{GSI Helmholtzzentrum f\"ur Schwerionenforschung GmbH, Planckstra{\ss}e 1, 64291 Darmstadt, Germany}
\address[c]{Physikalisches Institut, Ruprecht-Karls-Universit\"{a}t Heidelberg, 69120 Heidelberg, Germany}
\address[d]{National Research Nuclear University MEPhI, Moscow, Russia}
\address[e]{Physik Department E62, Technische Universit\"{a}t M\"{u}nchen, James-Franck-Str. 1, 85748, Garching, Germany}
\address[f]{Excellence Cluster Universe, Technische Universit\"{a}t M\"{u}nchen, Boltzmannstr. 2, 85748, Garching, Germany}
\address[g]{Pakistan Institute of Nuclear Science and Technology -- PINSTECH, Islamabad, Pakistan}
\address[i]{Institut für Kernphysik, Goethe-Universit\"{a}t, Max-von-Laue-Str.1, 60438 Frankfurt am Main, Germany}
\address[h]{Faculty of Mathematics, Physics and Informatics, Comenius University, Bratislava, Slovakia}

\input{abstract}


\end{frontmatter}


\input{introduction}

\input{experimental_set-up}

\input{anodePlaneAndHVprobeSignals}

\input{onsetAndTimeCurves}

\input{formationOfSecondaryDischarges}

\input{summary}

\newenvironment{acknowledgement}{\relax}{\relax}
\begin{acknowledgement}
\section*{Acknowledgements}
The Authors wish to thank the RD51 Collaboration, in particular F. Sauli, L. Ropelewski, E. Oliveri, and V. Peskov for fruitful discussions.\\ \indent
This research was supported by the DFG cluster of excellence ‘Origin and Structure of the Universe’ (www.universe-cluster.de); by the Federal Ministry of Education and Research (BMBF, Germany) [grant number 05P15WOCA1].
\end{acknowledgement}

\section*{References}
\bibliography{dc-paper}

\end{document}

%% file: abstract.tex
\begin{abstract}
Secondary discharges, which consist of the breakdown of a gap near a GEM foil upon a primary discharge across that GEM, are studied in this work.\\ \indent
Their main characteristics are the occurrence a few \SI{10}{\micro\second} after the primary, the relatively sharp onset at moderate electric fields across the gap, the absence of increased fields in the system, and their occurrence under both field directions.\\ \indent
They can be mitigated using series resistors in the high-voltage connection to the GEM electrode facing towards an anode. The electric field at which the onset of secondary discharges occurs indeed increases with increasing resistance. Discharge propagation form GEM to GEM in a multi-GEM system affects the occurrence probability of secondary discharges in the gaps between neighbouring GEMs.\\ \indent
Furthermore, evidence of charges flowing through the gap after the primary discharge are reported. Such currents may or may not lead to a secondary discharge. A characteristic charge, of the order of \SI{1e+10}{electrons}, has been measured as the threshold for a primary discharge to be followed by a secondary discharge, and this number slightly depends on the gas composition. A mechanism involving the heating of the cathode surface as trigger for secondary discharges is proposed.
\end{abstract}

%% file: introduction.tex
\section{Introduction}
\label{sec:intro}
Gas Electron Multiplier \cite{sauli1997gem} (GEM) foils are structures commonly used as charge amplification stages in proportional counters, due to their excellent performance at high particle rates. Being successfully employed in many experiments, \textit{e.g.} COMPASS \cite{Altunbas:02a}, LHCb \cite{Bencivenni:2002jr} or TOTEM \cite{Bagliesi:2010zz}, GEMs have become the solution for new experiments, such as the sPHENIX TPC \cite{Aidala:2012nz}, the upcoming upgrades of the CMS muon wall \cite{1748-0221-9-10-C10036} or the ALICE TPC \cite{aliceTpcUpgradeTDR2014}. In particular, the latter entails unprecedented challenges in terms of particle loads and performance goals.\\ \indent
GEMs are proven to work stably at particle rates of up to \SI{1}{\mega\hertz\per\centi\meter\squared} \cite{Altunbas:02a, Bencivenni:2002jr, Bagliesi:2010zz} or current loads exceeding \SI{10}{\nano\ampere\per\centi\meter\squared} \cite{Mathis:2018sjk}. However, the occurrence of high charge densities, possibly produced by the presence of highly ionising particles in the closest vicinity of the GEM foil, may significantly alter the stability of the detector. In turn, such high charge densities may trigger an electrical breakdown, which can result in damage of the foils or readout electronics. One of the key parameters for long-term operation of GEM-based detectors is therefore their stability against discharges.\\ \indent
The mechanism of a discharge in a GEM involves streamer development after reaching a critical amount of charges within a single GEM hole \cite{TUMsparks}. The overall stability of multi-GEM detectors can be, therefore, improved by proper optimisation of the high-voltage (HV) settings applied to the detector. The critical charge limits and the influence of HV settings on the stability of the GEM structures were thoroughly studied in \cite{Bachmann, TUMsparks, ALICEsparks, 1748-0221-12-05-C05017}.\\ \indent
\begin{figure*}
\centering
\includegraphics[width=0.75\textwidth, bb = 0 25 324 196, clip = true]{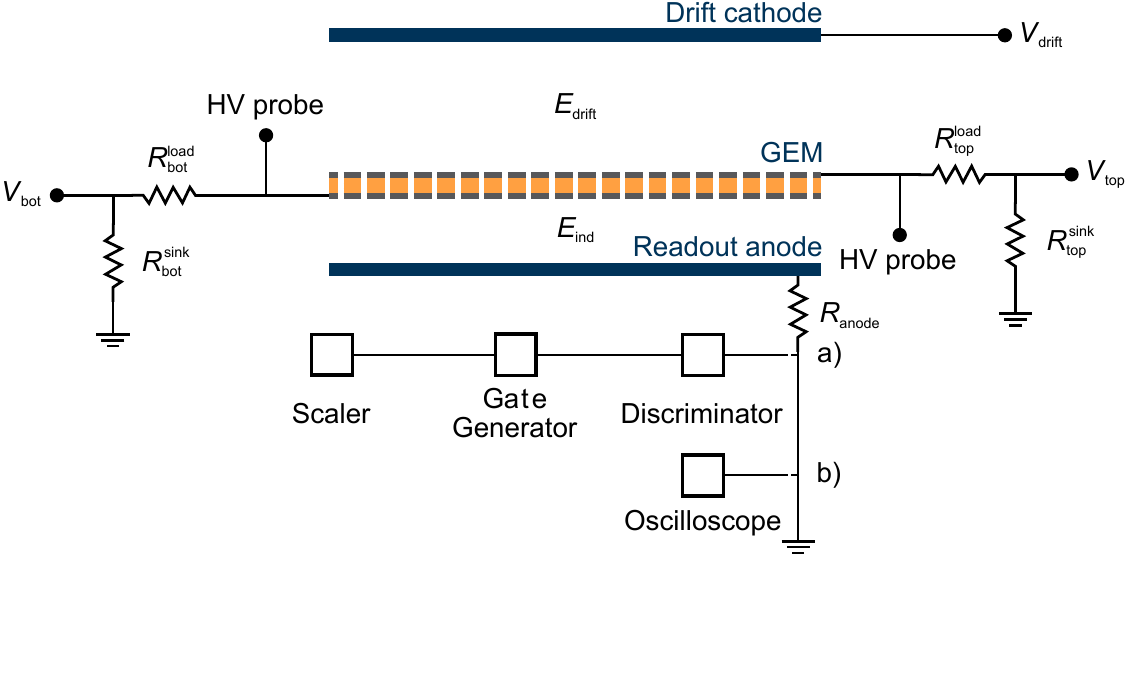}
\caption{\label{sec:setup:fig:setup} Schematic view of the GEM set-up: A single GEM is mounted on top of an anode with the cathode placed above the GEM. The discharges induce a signal on the anode plane, which is counted using NIM modules (a) and/or the corresponding waveform is recorded and stored with an oscilloscope (b). Two high-voltage probes are used at different electrodes in the detector to monitor and record the modification of the potentials during a discharge. The set-up can be extended by another GEM, which is not shown here. In the double-GEM mode we label the GEM facing the drift volume GEM1, while the GEM below is referred to as GEM2.}
\end{figure*}
It was, however, observed in \cite{Bachmann, Peskov} that primary discharges in GEM holes may trigger a secondary discharge, which occurs with a delay of up to several ten microseconds below the discharging GEM.\footnote{When using above (below, respectively) a GEM we refer to the gap in the vicinity of the GEM with lower or more negative (higher or more positive, respectively) electrical potential.} Secondary discharges may occur in single and multi GEM stacks between adjacent foils and/or between the last foil in the stack and the readout anode. Due to the large amount of energy carried by this kind of discharges, they may cause fatal damage to the amplification structure. As shown in \cite{Bachmann}, the secondary discharge probability, defined as the ratio of the observed secondary discharges to the total GEM discharges, increases with the strength of the electric field between subsequent GEM foils and with the energy of the primary discharge. Additionally, the secondary discharges appear already at fields lower than necessary for amplification in the gas.\\ \indent
The underlying mechanism of the formation of secondary discharges is unknown. Various possible explanations for their occurrence have been proposed over the years \cite{Peskov}, including the sudden increase of the field below the discharging GEM.\\ \indent
In this work, we provide further insights into the nature of secondary discharges and we present a method to mitigate their occurrence. The paper is organised as follows: in \secref{sec:setup} the experimental set-up is described. Different measurements of secondary discharge characteristics are presented in \secref{sec:results:signalCharacteristics}. These include studies to determine whether a secondary discharge originates in the counting gas or if it results from the RC-characteristics of the detector and the HV system (\secrefbra{sec:results:subsec:GDTstudies}). In \secref{sec:results:onsetcurves} measurements exploring the dependence of the secondary discharge probability on various parameters are presented. We study secondary discharges for different field configurations and directions. Furthermore, we present a way of mitigating the occurrence of secondary discharges (\secrefbra{sec:results:mitigationWithDecouplingResistors}) and explain how this mitigation works (\secrefbra{sec:results:onsetcurves:subsec:decouplingR}). In \secref{sec:formationSecDCs} we propose a production mechanism for secondary discharges, based on our findings presented in this paper. Finally, we summarise and conclude our findings in \secref{sec:summary}.

%% file: experimental_set-up.tex
\section{Experimental set-up}
\label{sec:setup}
\Figref{sec:setup:fig:setup} shows a schematic view of the experimental set-up and its powering scheme. The detector vessel contains either one or two $10\times\SI{10}{\centi\meter\squared}$ GEM foils with a readout anode below and a cathode above the GEMs. The gap between the two GEMs (transfer gap) and the gap between the readout anode and the closest GEM (induction gap) is fixed to \SI{2}{\milli\meter}. We employ an open gas system with mass-flow meters that allows to mix up to three gases. The detector volume is flushed with either \arco{} with different admixtures of $\textrm{CO}_2$ or \baseline{}, where the numbers in the bracket represent the mixing ratio. For this work, either a mixed $^{239}\textrm{Pu}+\ ^{241}\textrm{Am}+\ ^{244}\textrm{Cm}$ source mounted in a fixed position inside the gas volume or $^{222}\textrm{Rn}$, which is released into the gas stream, are used to induce discharges. The GEM foils, produced at the CERN PCB workshop with the double-mask technology, are of standard design: \SI{50}{\micro\meter} thick polyimide (Apical) covered on both sides with a \SI{5}{\micro\meter} copper layer, perforated with holes with \SI{50}{\micro\meter} inner and \SI{70}{\micro\meter} outer hole diameter at a pitch of \SI{140}{\micro\meter}.\\ \indent
Potentials are applied to the cathode ($V_{\textrm{drift}}$) and each GEM electrode via independent channel power supplies. The GEM potentials $V_{\textrm{top}}$ and $V_{\textrm{bot}}$ at the top and bottom electrode of the GEM are applied via the loading resistor $R^{\textrm{load}}_{\textrm{top}}$ and the decoupling resistor $R^{\textrm{load}}_{\textrm{bot}}$. The voltage difference $\Delta V_{\textrm{GEM}} = V_{\textrm{top}} - V_{\textrm{bot}}$ is adjusted to the end of observing a discharge rate on the order of \SI{1}{\hertz}. When we use two GEMs, the voltage difference across the GEM closest to the drift region (GEM1), $\Delta V_{\textrm{GEM1}}$, is chosen sufficiently low so the GEM does not discharges while amplifying primary ionisations. The voltage difference of the GEM closer to the readout anode (GEM2), $\Delta V_{\textrm{GEM2}}$, is then adjusted to reach a moderate rate of GEM2 discharges.\\ \indent
The readout anode is connected to ground via the $R_{\textrm{anode}}$ resistor. Its resistance is in the order of \SI{10}{\ohm} as given by the attenuator used to attenuate the signal. An attenuating circuit is mandatory in order to protect the measurement devices and because the non-attenuated discharge signals can saturate the dynamic range of the oscilloscopes used. For dedicated measurements either $R_{\textrm{anode}}$ in the \si{\kilo\ohm} range is used or the readout anode is biased with a positive potential.\\ \indent
In case of a discharge in a GEM the loading resistor $R^{\textrm{load}}_{\textrm{top}}$ quenches the current induced by the discharge. Its value is set to either \SI{5}{\mega\ohm} or \SI{10}{\mega\ohm}. $R^{\textrm{load}}_{\textrm{bot}}$ is varied to investigate its effect on the secondary discharge probability. To ensure a safe and fast discharge of the GEM after a power supply trip, resistors to ground $R_{\textrm{top}}^{\textrm{sink}}$ and $R_{\textrm{bot}}^{\textrm{sink}}$ are installed. However, some measurements are performed without having this path to ground and during these measurements there are as well no $R^{\textrm{sink}}$ resistors. We indicate this detector configuration with \textit{No $R^{\textrm{sink}}$} as we present the corresponding measurements.\\ \indent
$V_{\textrm{drift}}$ and $V_{\textrm{top}}$ define the drift field $E_{\textrm{drift}}$ above the GEM, and the potential difference between $V_{\textrm{bot}}$ and the anode defines the induction field $E_{\textrm{ind}}$. A drift field of \SI{400}{\volt\per\centi\meter} is used for the bulk of measurements presented in this work. During measurements we vary either the induction field or the transfer field.\\ \indent
\label{sec:setup:subsec:detectorReadout}
In order to directly measure the potentials on the GEM electrodes during and after a discharge, HV probes are connected to the GEM electrodes. We use two custom made probes which contain each a resistance of \SI{340}{\mega\ohm} in parallel to a series of 22 capacitors, each \SI{1.3}{\pico\farad}. With this configuration the probes' resistance and capacitance (RC) match the input RC value of a digital oscilloscope, to which the probes are connected. Thus, the probes' impedance together with the input impedance of the oscilloscope act as a voltage divider and allow to show the scaled-down potential at an electrode of interest.\\ \indent
The anode is read out via two signal branches after the signal is attenuated (see \figrefbra{sec:setup:fig:setup}). In branch a), the signal is processed by a chain of a discriminator, a gate generator, and a scaler. The counting logic counts and differentiates primary and secondary discharges. In branch b), the signal is fed to the oscilloscope to record the waveforms together with the HV probe signals.

%% file: anodePlaneAndHVprobeSignals.tex
\section{Discharge signal characteristics}
\label{sec:results:signalCharacteristics}
In this section characteristic features of primary and secondary discharge signals are discussed, using data recorded with the oscilloscope. To this end, we use individual waveforms to illustrate features, which can be observed in every recording. We first consider the case of primary discharges occurring in the GEM closest to the readout anode and subsequent secondary discharges occurring in the induction gap. \Secref{sec:results:subsec:2ndarysInTGap} covers the corresponding findings for secondary discharges in a gap between two GEMs.

\subsection{Primary and secondary discharges}
\label{sec:results:subsec:primaryAndSecondary}
\begin{figure}
\centering
\includegraphics[width=0.49\textwidth, bb = 0 0 720 538, clip = true]{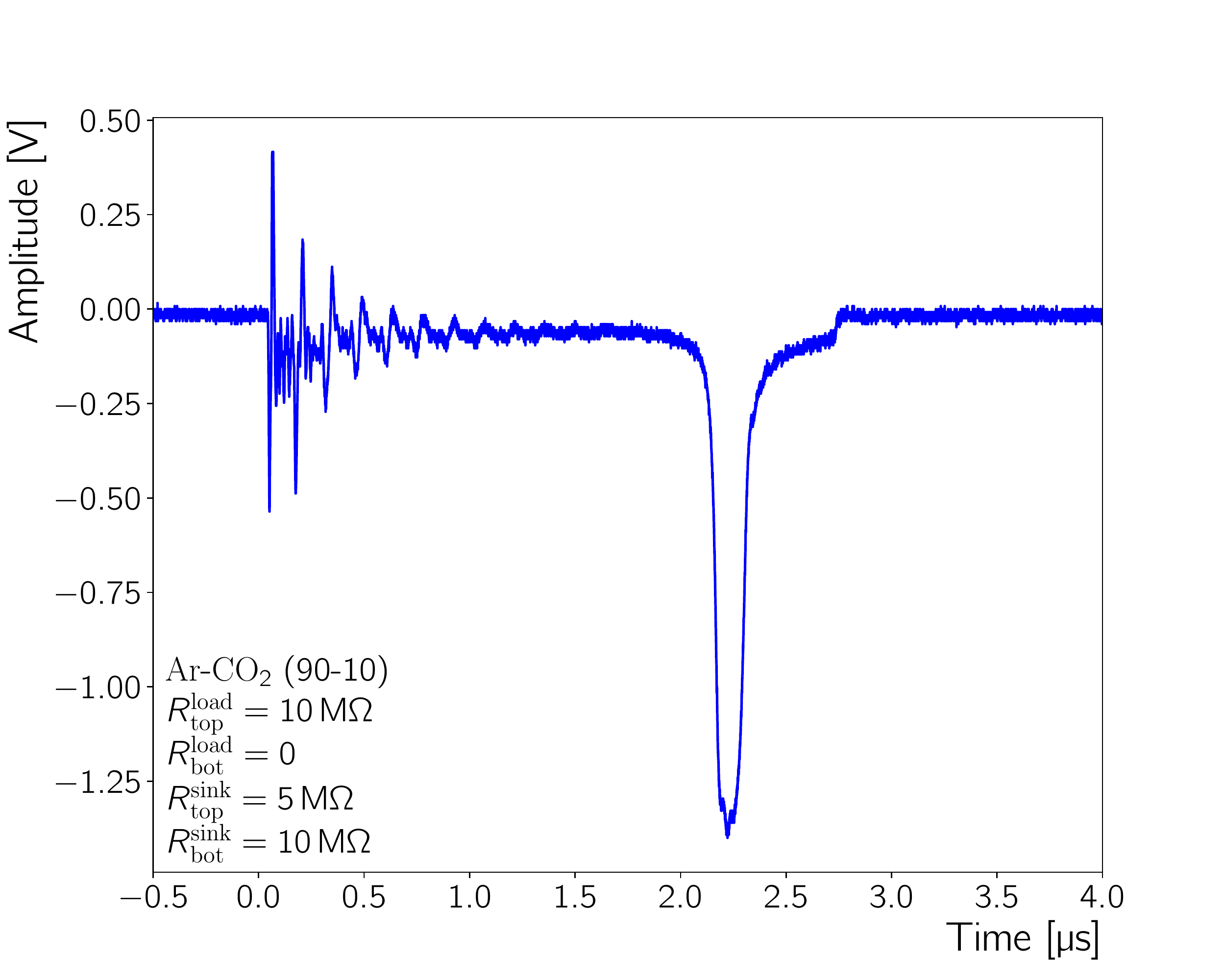}
\caption{\label{sec:results:fig:readoutPlaneSignalsLowResistanceToGND:sec}Readout anode signal of a discharge followed by a secondary discharge. The waveform is recorded with a single GEM set-up. After the readout anode the signal is passed through a \SI{32}{dB} attenuator, which provides $R_{\textrm{anode}}$ of about \SI{10}{\ohm}.}
\end{figure}

\Figref{sec:results:fig:readoutPlaneSignalsLowResistanceToGND:sec} shows at $t\sim\,0$ the temporal evolution of a primary discharge in a GEM recorded on the anode; primary discharges appear as a fast oscillating signal around the baseline when using an attenuating circuit with a $R_{\textrm{anode}}$ of a few \si{\ohm}. For increasing $R_{\textrm{anode}}$ the signal is more and more dominated by the charge deposited on the readout anode and the time needed to discharge it again. Furthermore the oscillation frequency and duration changes according to the (parasitic) capacitances in the system and the value of $R_{\textrm{anode}}$.\\ \indent
\begin{figure*}
\centering
\subfloat[]{\includegraphics[width=0.49\textwidth, bb = 0 0 720 538, clip = true]{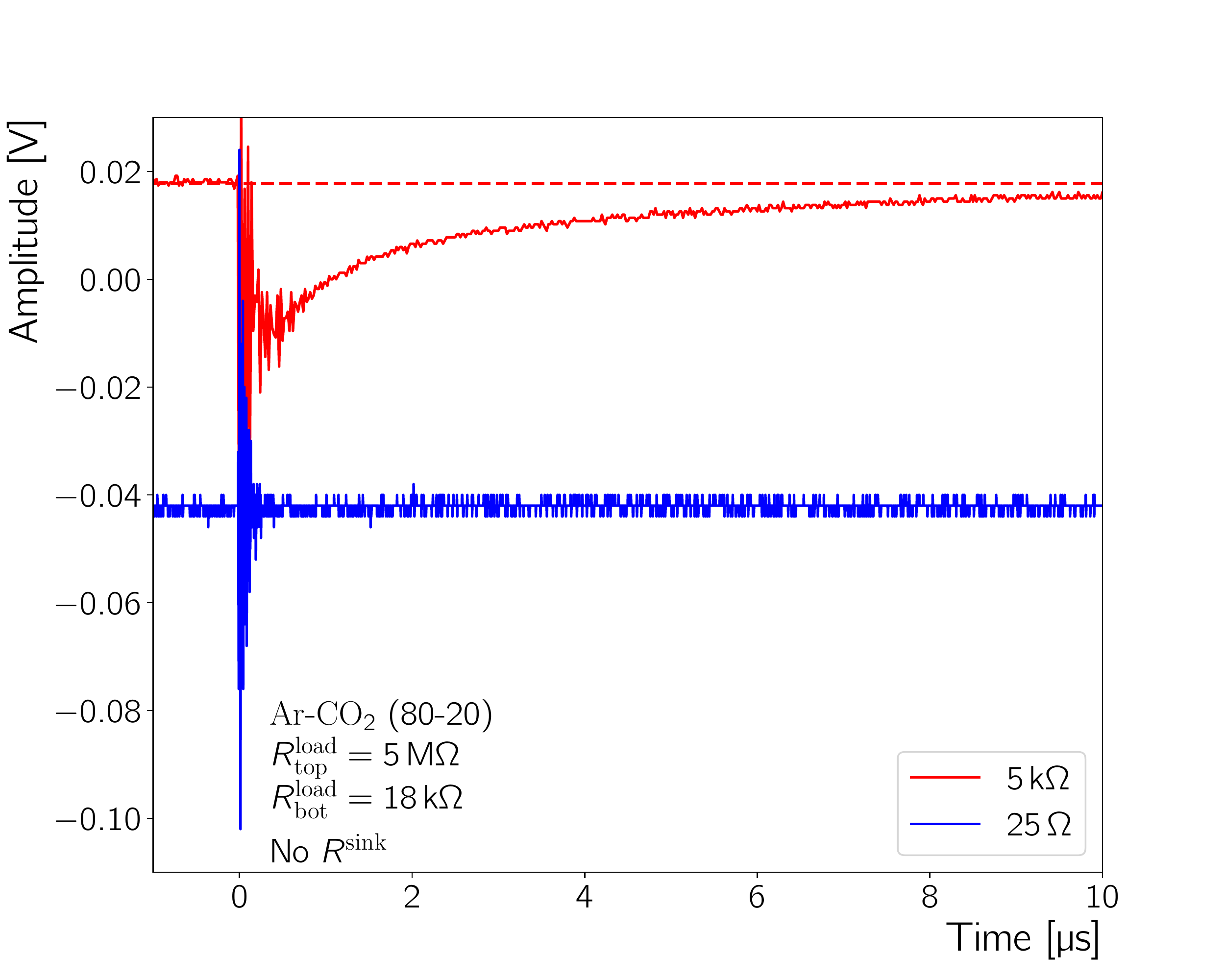}
\label{sec:results:fig:readoutPlaneSignalsHighAndLowResistanceToGND:noSec}}
\subfloat[]{\includegraphics[width=0.49\textwidth, bb = 0 0 720 538, clip = true]{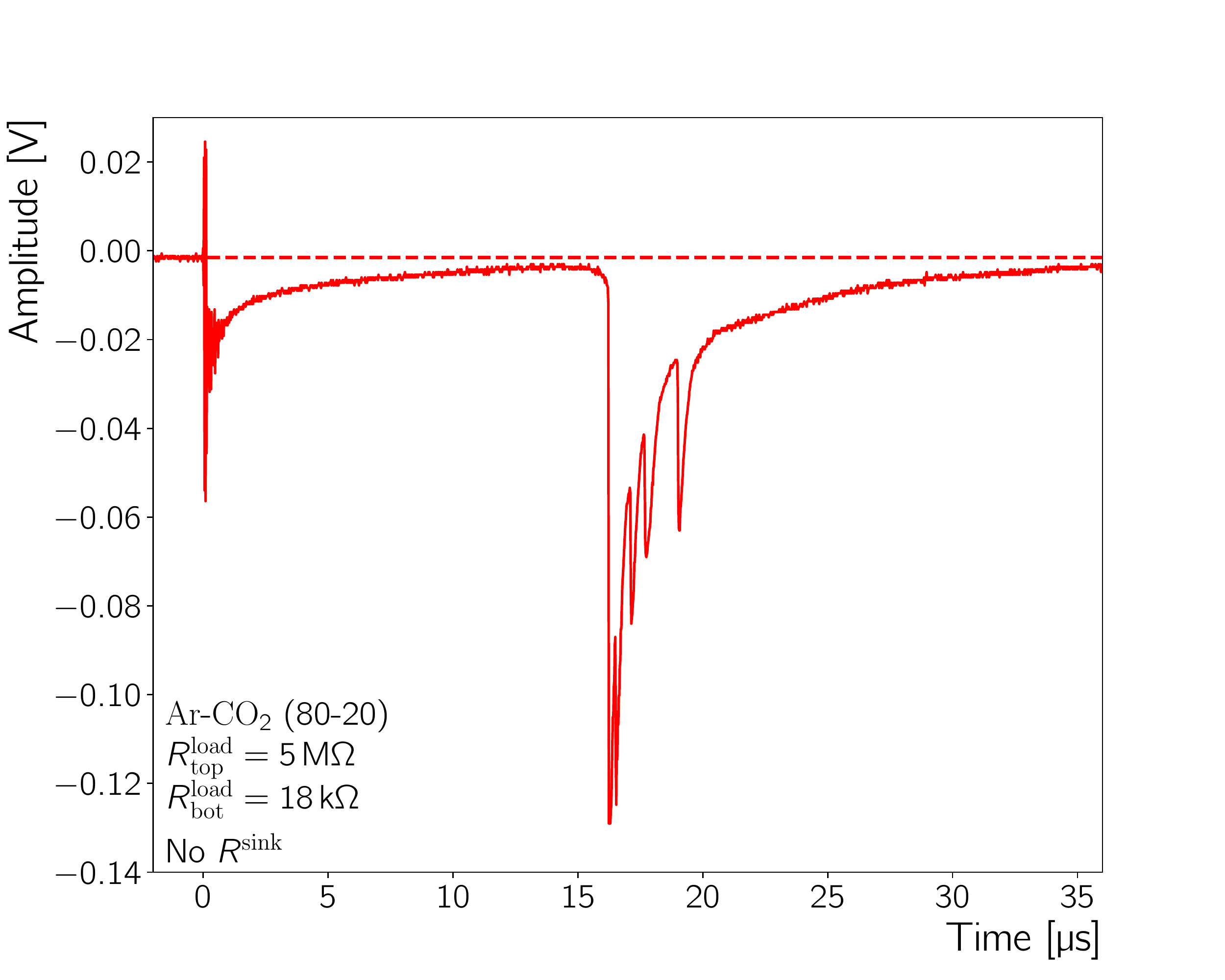}
\label{sec:results:fig:readoutPlaneSignalsHighAndLowResistanceToGND:sec}}
\caption{\label{sec:results:fig:readoutPlaneSignalsHighAndLowResistanceToGND:noSecAndSec}(Colour online.) \protect\subref{sec:results:fig:readoutPlaneSignalsHighAndLowResistanceToGND:noSec} Readout anode signal of a discharge for $R_{\textrm{anode}}=\SI{25}{\ohm}$, blue,  and \SI{5}{\kilo\ohm}, red. Both signals have been shifted on the vertical axis in order to improve readability. \protect\subref{sec:results:fig:readoutPlaneSignalsHighAndLowResistanceToGND:sec} An initial and secondary discharge ($t\sim\,\SI{16}{\micro\second}$) recorded on the readout anode during measurements with $R_{\textrm{anode}}=\SI{5}{\kilo\ohm}$. In both plots a dashed line indicates the baseline of the signal recorded with $R_{\textrm{anode}}=\SI{5}{\kilo\ohm}$.}
\end{figure*}
Around \SI{2.2}{\micro\second} after the initial signal a secondary discharge is observed, seen as a large negative pulse (see \figrefbra{sec:results:fig:readoutPlaneSignalsLowResistanceToGND:sec}). The signal shape of a secondary discharge depends on the value of $R_{\textrm{anode}}$ and other circuit elements, as it is the case for the primary discharge signal shape. The inspection of all recorded waveforms shows that a secondary discharge always follows a primary discharge and never occurs alone.

\subsection{Currents preceding the secondary discharge}
\label{sec:results:subsec:currentsBefore2NDary}
A high resistance in the readout anode's connection to ground allows to study currents through the induction gap. While for a resistance to ground of \SI{25}{\ohm} only an oscillating signal amplitude is visible (\figrefbra{sec:results:fig:readoutPlaneSignalsHighAndLowResistanceToGND:noSec}, blue) these oscillations sit on top of a distinct unipolar signal when this resistance is increased to \SI{5}{\kilo\ohm}, as shown in \figref{sec:results:fig:readoutPlaneSignalsHighAndLowResistanceToGND:noSec} (red).\\ \indent
This is a strong indication for a current through the induction gap, which decays on a time scale longer than \SI{10}{\micro\second}. However, in case of a secondary discharge, the current increases again as can be seen in \figref{sec:results:fig:readoutPlaneSignalsHighAndLowResistanceToGND:sec}: the readout voltage starts to drop slightly, approximately \SI{1}{\micro\second} before the secondary discharge is visible, as seen at $t\sim\,\SI{15}{\micro\second}$ in the figure. The secondary discharge itself ($t\sim\,\SI{16}{\micro\second}$) is characterised by a large, almost instantaneous voltage breakdown in the induction gap, which is visible as a fast change of the signal amplitude.\\ \indent
The first slight drop between \SI{15}{\micro\second} and \SI{16}{\micro\second}, where the current stops decaying and increases again, is a feature pointing to a preparatory mechanism preceding the secondary discharge.

\subsection{Evolution of GEM potentials}
\label{sec:results:subsec:HVProbeSignals}
\begin{figure*}
\centering
\subfloat[]{
\includegraphics[width=0.49\textwidth, bb = 0 0 720 538, clip = true]{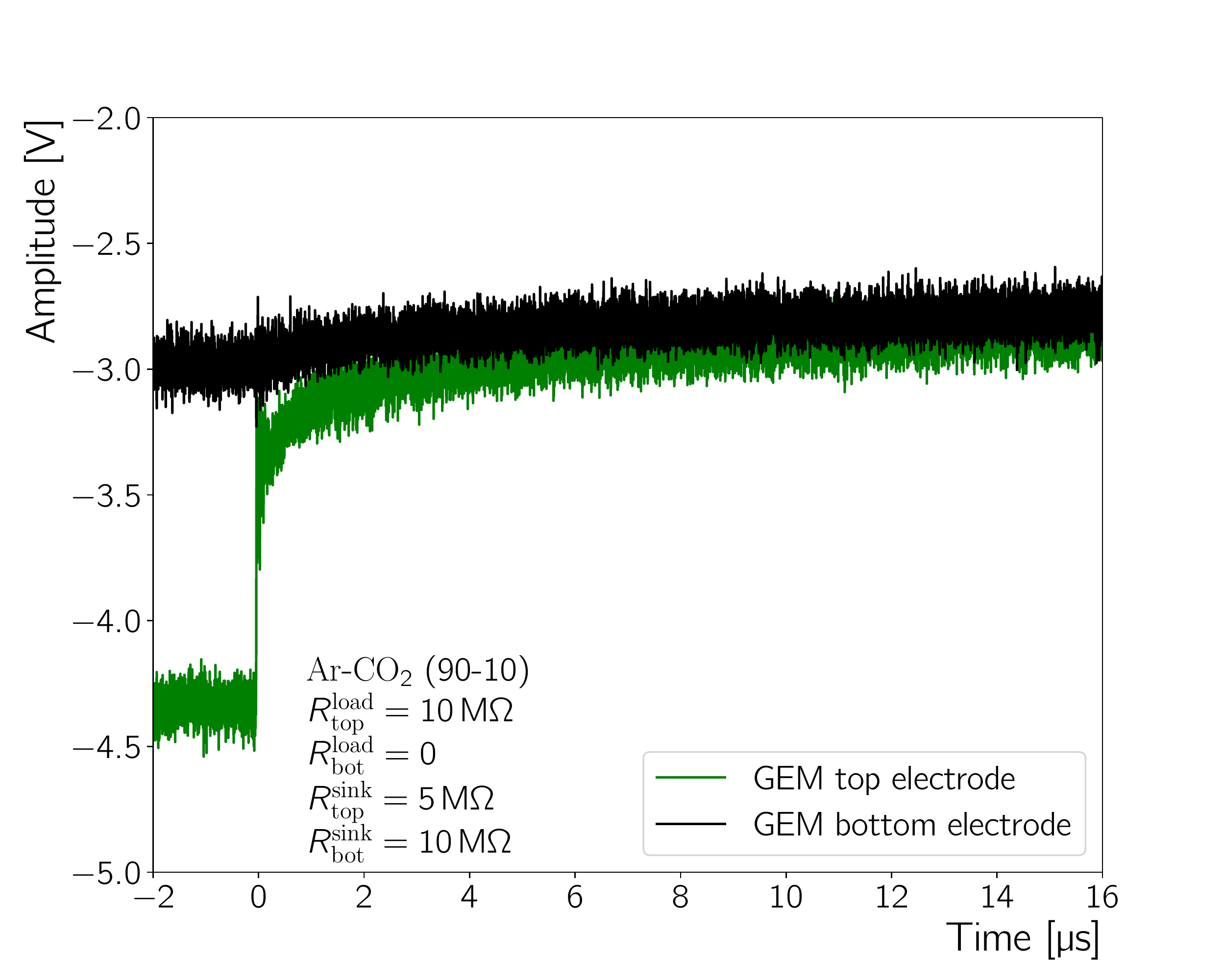}\label{sec:results:fig:HVprobeSignals:SingleGEM:noSec}}
\subfloat[]{
\includegraphics[width=0.49\textwidth, bb = 0 0 720 538, clip = true]{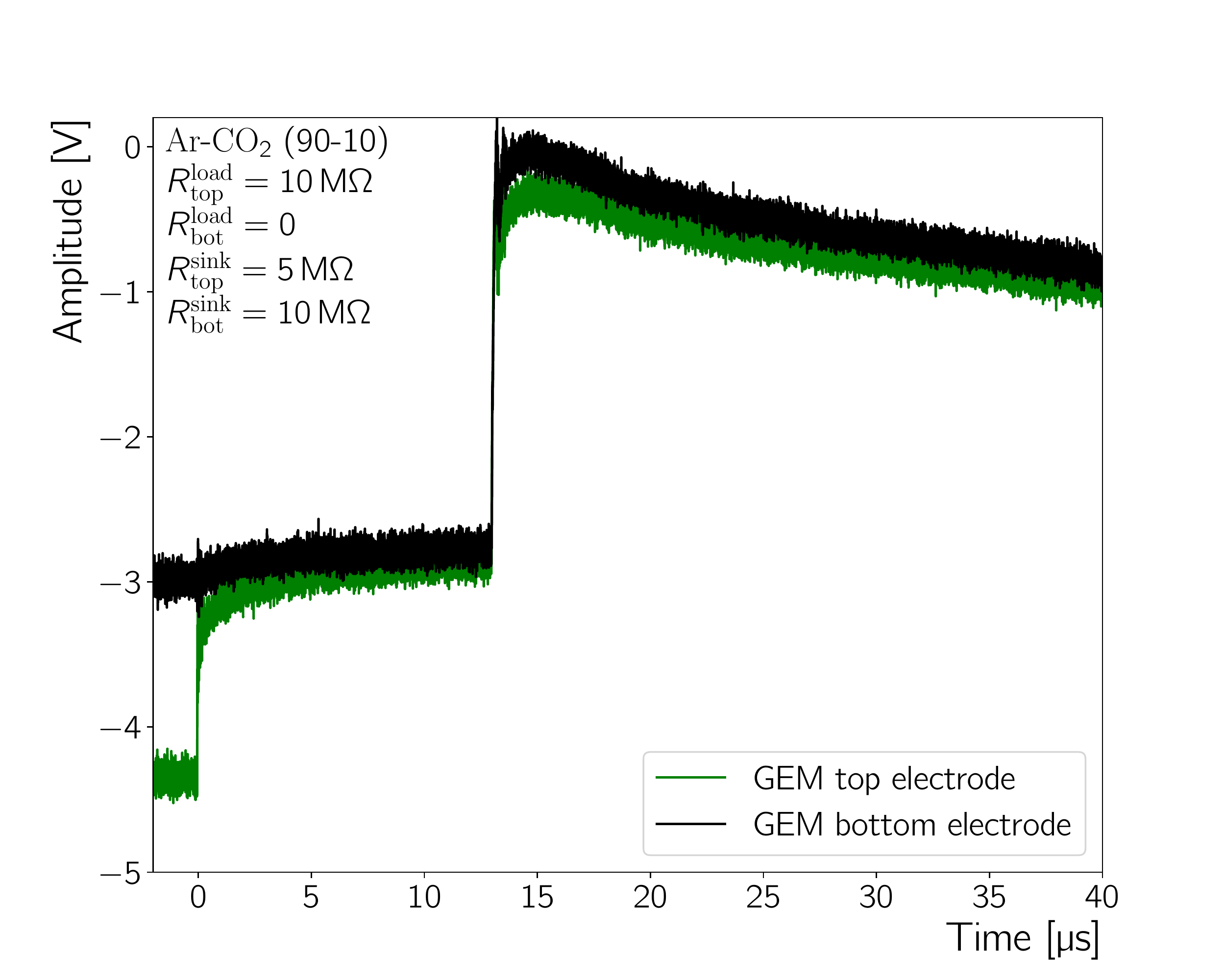}\label{sec:results:fig:HVprobeSignals:SingleGEM:sec}}
\caption{\label{sec:results:fig:HVprobeSignals:SingleGEM:noSecAndSec}(Colour online.)  \protect\subref{sec:results:fig:HVprobeSignals:SingleGEM:noSec} Potentials at the top and bottom electrodes of a GEM during a primary discharge and \protect\subref{sec:results:fig:HVprobeSignals:SingleGEM:sec} during another primary discharge, followed by a secondary discharge. These waveforms have been recorded with HV probes at an induction field of \SI{5}{\kilo\volt\per\centi\meter}.}
\end{figure*}
As a next step the potentials of the GEM electrodes are recorded with HV probes (see \secrefbra{sec:setup:subsec:detectorReadout} for details) in addition to the signal recorded at the readout anode. In order to measure the potential directly at the GEM electrode, the probes are connected to the respective GEM electrode.\\ \indent
The primary discharge is formed across the electrodes of a GEM, it discharges the capacitor formed by these electrodes. Consequently, the voltage difference between the bottom and top GEM electrode drops as the GEM discharges ($t=0$ in \figrefbra{sec:results:fig:HVprobeSignals:SingleGEM:noSecAndSec}). This voltage drop affects mainly the potential at the top GEM electrode, as $R_{\textrm{top}}^{\textrm{load}}\gg R_{\textrm{bot}}^{\textrm{load}}$, while the potential at the bottom electrode of the GEM is approximately constant, since $R_{\textrm{bot}}^{\textrm{load}}=0$. Thus, at the time of the secondary discharge $\Delta V_{\textrm{GEM}}$ is practically zero. A strong voltage change occurs during a secondary discharge ($t\sim\,\SI{14}{\micro\second}$ in \figrefbra{sec:results:fig:HVprobeSignals:SingleGEM:sec}), where the voltage approaches the readout anode potential, which is at ground. The observation that $V_{\textrm{bot}}$ and $V_{\textrm{top}}$, which are coupled by the capacity of the GEM, approach the potential of the anode plane during the secondary discharge and vice versa suggests that a secondary discharge corresponds to the breakdown in the gas gap between GEM and readout anode. The potential difference across the GEM is not significantly altered before or during the secondary discharge. Furthermore, there is no increase of the electric field in the induction gap or anywhere in the system, which could explain the occurrence of this breakdown phenomenon.

\subsection{Photographs of a primary and a secondary discharge}
\label{sec:results:subsec:dischargePhotos}
\begin{figure}
\centering
\subfloat[]{\includegraphics[width=0.49\textwidth, bb = 0 0 320 60, clip = true]{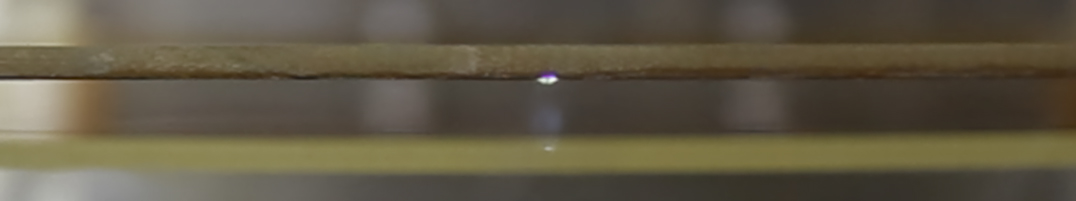}\label{sec:results:fig:primaryDischargePhoto}}\\
\subfloat[]{\includegraphics[width=0.49\textwidth, bb = 0 0 320 60, clip = true]{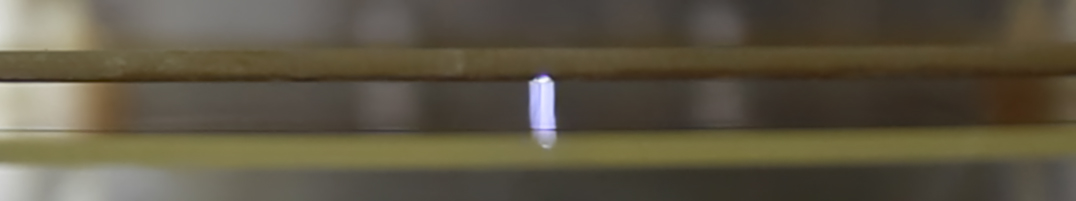}\label{sec:results:fig:secondaryDischargePhoto}}
\caption{\label{sec:results:fig:dischargePhoto}Side view on the induction gap. In both photographs the GEM (upper structure) and the readout anode (lower structure) are visible. Photo \protect\subref{sec:results:fig:primaryDischargePhoto} shows a primary discharge while photo \protect\subref{sec:results:fig:secondaryDischargePhoto} shows a secondary discharge between GEM and readout anode.}
\end{figure}
The photographs in \figref{sec:results:fig:dischargePhoto} confirm our conclusion that the secondary discharge corresponds to the breakdown in a gap. A single GEM set-up is placed in a transparent gas vessel and filmed with a CANON 5D Mark III camera. Afterwards, the film is analysed frame by frame and the frames with discharges are extracted. \Figref{sec:results:fig:primaryDischargePhoto} shows a frame containing a primary discharge, while a secondary discharge is visible in the frame displayed in \figref{sec:results:fig:secondaryDischargePhoto}. The arc in the latter figure depicts the light emission of a secondary discharge, which is triggered by a current flow and voltage breakdown in the induction gap. The secondary discharge can thus clearly be related to the breakdown of the induction gap. This confirms the conclusions drawn from the HV probe measurements in \secref{sec:results:subsec:HVProbeSignals}.

\subsection{Discharge studies with Gas Discharge Tubes}
\label{sec:results:subsec:GDTstudies}
With the observation of secondary discharges the following question arises: are the currents observed before a secondary discharge (see \secrefbra{sec:results:subsec:currentsBefore2NDary}) and the features seen in the readout anode and HV probe signals (see \secs\ref{sec:results:subsec:currentsBefore2NDary} and \ref{sec:results:subsec:HVProbeSignals}) effects in the counting gas or are they a result of the response of the circuit elements and the power supply to the primary discharge? In order to answer this question, tests with Gas Discharge Tubes (GDTs) are carried out. A GDT is a commercially available component which consists of a gas encapsulated in a cylinder with electrodes on the sides \cite{GDTs}. They are designed to discharge as soon as the applied voltage exceeds their nominal breakdown voltage ($V_{\textrm{GDT}}$). Therefore, GDTs allow to induce discharges in a circuit, and thus to test the circuits response to discharges without the effects of moving charges in the counting gas.\\ \indent
\begin{figure*}
\centering 
\subfloat[]{\includegraphics[width=0.24375\textwidth, bb = 0 200 220 405, clip = true]{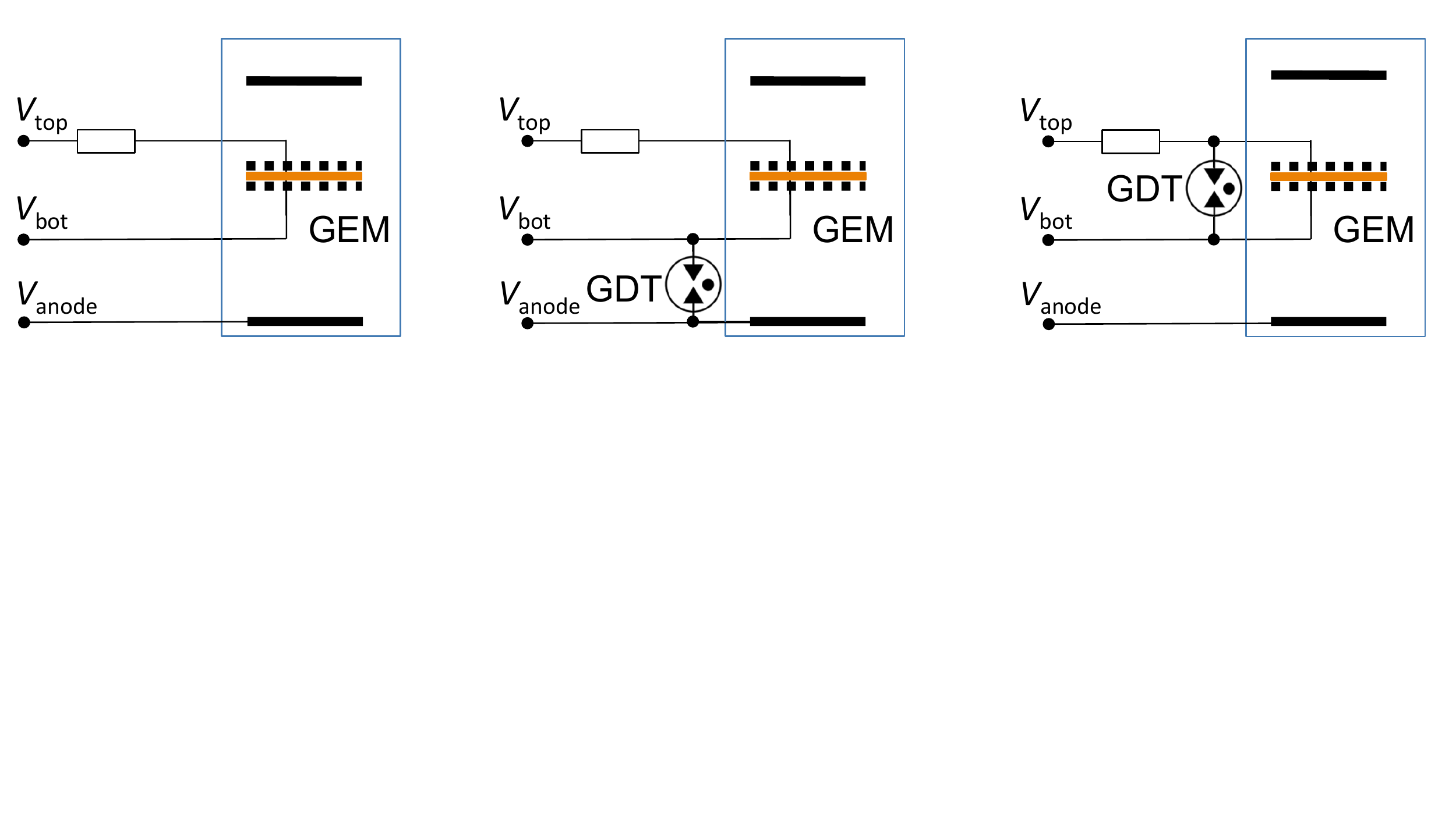}\label{sec:results:fig:GDTsetups:a}}\subfloat[]{\includegraphics[width=0.24375\textwidth, bb = 240 200 470 405, clip = true]{GDT_set-ups.pdf}\label{sec:results:fig:GDTsetups:b}}\subfloat[]{\includegraphics[width=0.24375\textwidth, bb = 500 200 720 405, clip = true]{GDT_set-ups.pdf}\label{sec:results:fig:GDTsetups:c}}
\caption{\label{sec:results:fig:GDTsetups}Sketch of different configurations of the detector used with Gas Discharge Tubes: \protect\subref{sec:results:fig:GDTsetups:a} The detector is present in the standard configuration without any GDT. \protect\subref{sec:results:fig:GDTsetups:b} A GDT is mounted between the GEM bottom electrode and the readout anode. \protect\subref{sec:results:fig:GDTsetups:c} One GDT is present in parallel to the GEM, it is thus in contact with the GEM bottom and top electrodes. Two HV configurations are used for these detector configurations: I) $V_{\textrm{anode}}$ equals ground potential and the other electrodes are supplied with a negative voltage. II) $V_{\textrm{anode}}$ is a positive voltage, $V_{\textrm{bot}}$ at ground potential and $V_{\textrm{top}}$ at a negative potential. Some details depicted in \figref{sec:setup:fig:setup} are omitted here.}
\end{figure*}
GDTs are added to the experimental set-up as shown in \figref{sec:results:fig:GDTsetups}. The detector's standard configuration is shown in Figure~\ref{sec:results:fig:GDTsetups:a}. When $V_{\textrm{anode}}$ is at a postivie voltage, the GEM bottom electrode is solidly grounded. This configuration is tested to ensure that $V_{\textrm{bot}}$ does not change during the discharge processes. The field strength of the induction field, however, is maintained at the same value as when the readout anode is at ground potential. For both voltage settings secondary discharges are observed at the same induction field. Systematic studies of this field dependence will be presented in \secref{sec:results:onsetcurves}.\\ \indent
\begin{figure*}
\centering
\subfloat[]{
\includegraphics[width=0.49\textwidth, bb = 0 0 720 538, clip = true]{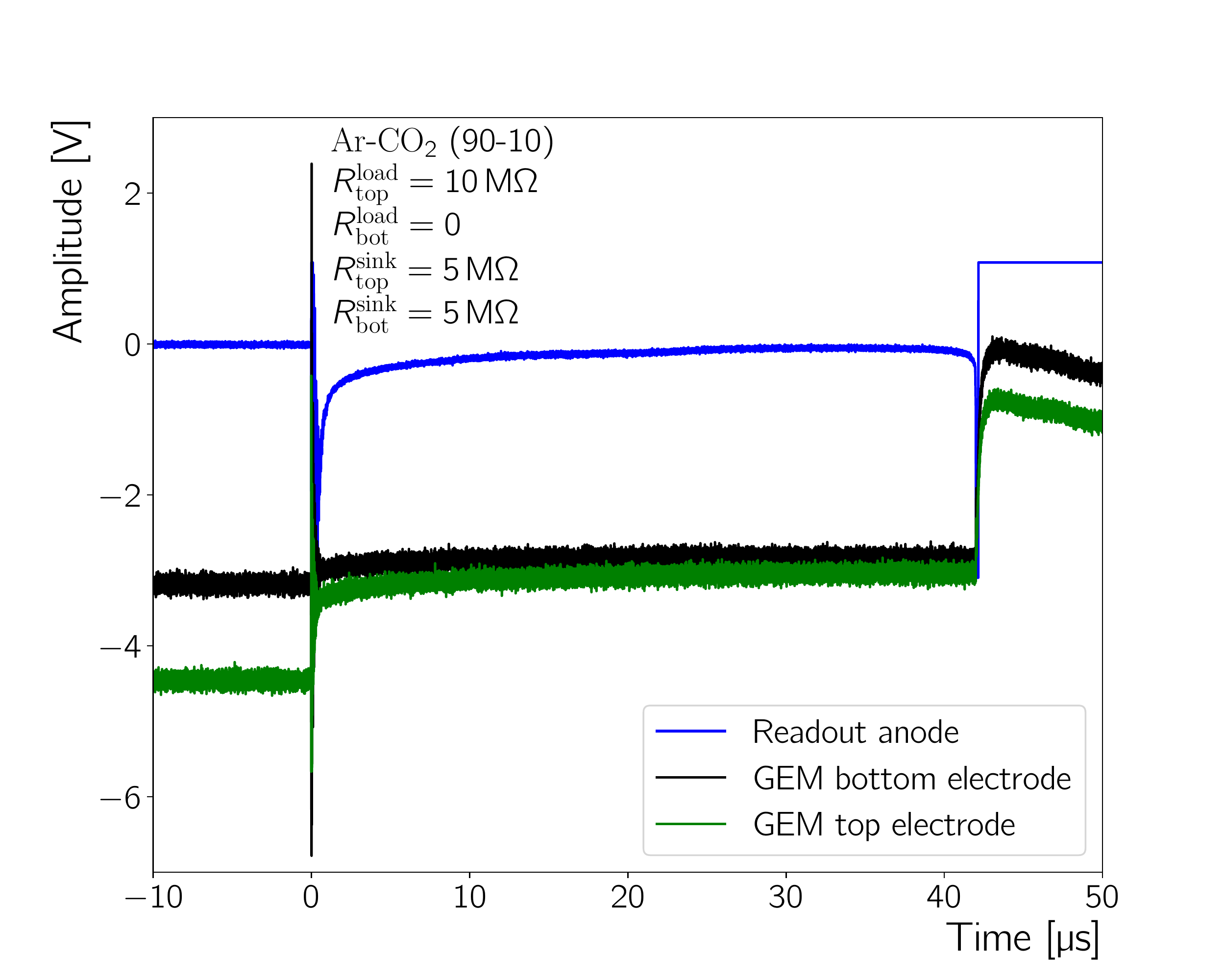}\label{sec:results:fig:HVprobeSignals:GDTs:config1HV}}
\subfloat[]{
\includegraphics[width=0.49\textwidth, bb = 0 0 720 538, clip = true]{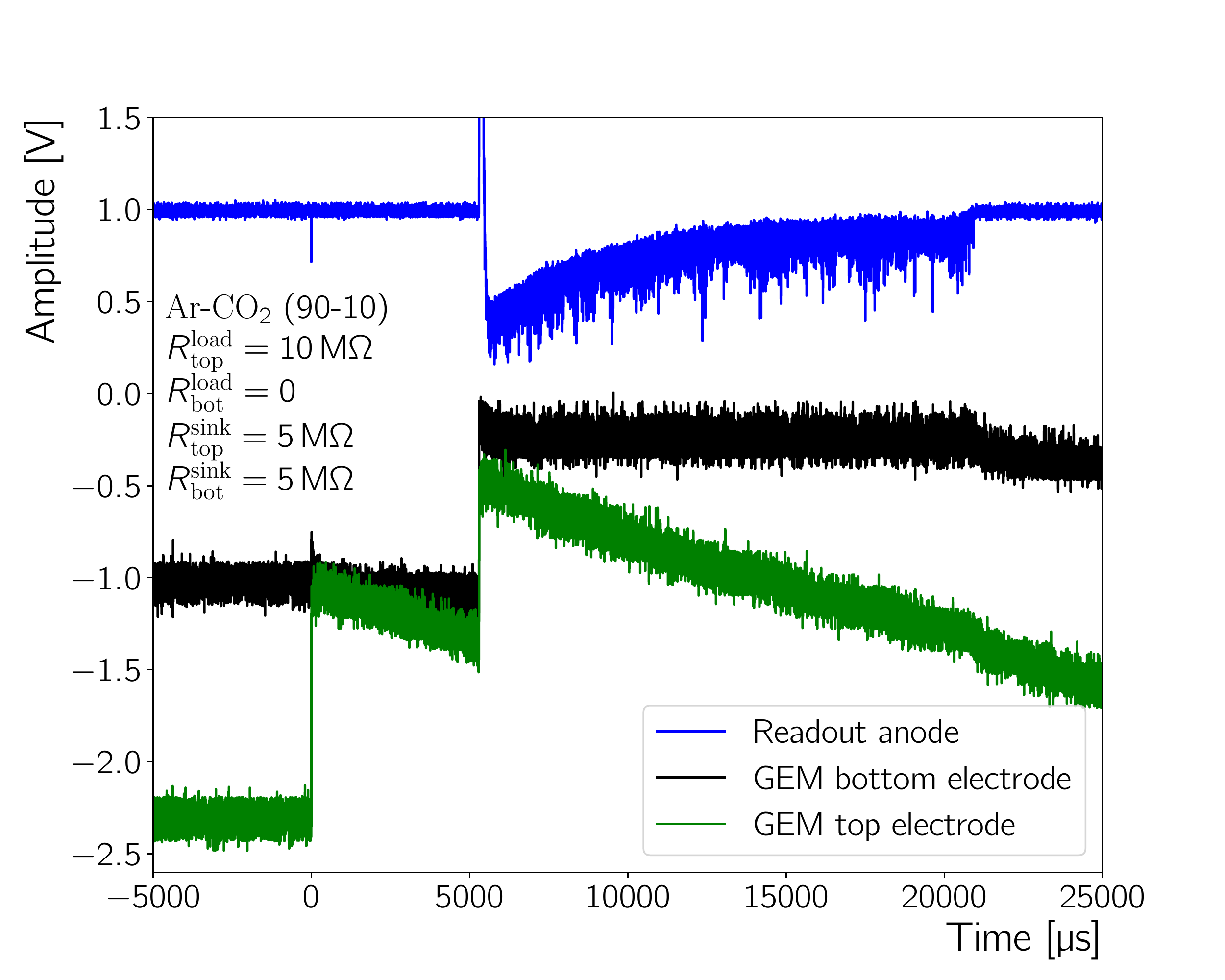}\label{sec:results:fig:HVprobeSignals:GDTs:config2HV}}
\caption{\label{sec:results:fig:HVprobeSignals:GDTs:config2}(Colour online) Waveforms measured for a discharge in a GEM followed by \protect\subref{sec:results:fig:HVprobeSignals:GDTs:config1HV} a secondary discharge in the induction gap in case of the standard detector configuration (Fig. \ref{sec:results:fig:GDTsetups:a}) or \protect\subref{sec:results:fig:HVprobeSignals:GDTs:config2HV} a discharge in a Gas Discharge Tube mounted in parallel to the induction gap (Fig. \ref{sec:results:fig:GDTsetups:b}). In figure \protect\subref{sec:results:fig:HVprobeSignals:GDTs:config2HV} the readout anode signal has been vertically shifted for better visibility. Note the different time scale on which the GDT discharges occur. Furthermore, an increase of the GEM bottom electrode potential for $0<t<\SI{5000}{\micro\second}$ is visible, causing the GDT discharge.}
\end{figure*}
A typical event in the standard configuration is displayed in \figref{sec:results:fig:HVprobeSignals:GDTs:config1HV}, including a primary and a secondary discharge. The features discussed in \secrefs{sec:results:subsec:currentsBefore2NDary} and \ref{sec:results:subsec:HVProbeSignals} are well visible. For comparison, an event recorded with a GDT mounted in parallel to the induction gap (\figrefbra{sec:results:fig:GDTsetups:b}) is displayed in \figref{sec:results:fig:HVprobeSignals:GDTs:config2HV}. The potential difference across the induction gap is adjusted to be slightly below the $V_{\textrm{GDT}}$, which in turn is sufficiently lower than the threshold for the occurrence of secondary discharges in the induction gap. However, upon a primary discharge, a slight increase (on a long time scale on the order of \si{\milli\second}) of the GEM bottom electrode potential is observed (\figrefbra{sec:results:fig:HVprobeSignals:GDTs:config2HV}, black line) which eventually leads to the discharge of the GDT. In addition, no current in advance of the GDT discharge is observed, which also shows that the GDT discharge is not comparable to a secondary discharge.\\ \indent
To avoid the increase of the GEM bottom potential, the corresponding GEM electrode is connected to ground and the readout anode is biased with positive HV as discussed before. No increase of the GEM bottom potential, and thus no discharges of the GDT, is observed while biasing the detector in this way, proving that the previously observed GDT discharges are not caused by the discharge itself, but by the reaction of the circuit to the discharge.\\ \indent
For the third series of tests the GDT is connected in parallel to the GEM (\figrefbra{sec:results:fig:GDTsetups:c}). A GDT with $V_{\textrm{GDT}}$ smaller than the voltage for a discharge in the GEM is used, ensuring that only the GDT discharges, but not the GEM. Doing so, the full biasing circuit still has to respond to the discharge, but the discharge itself is decoupled from the counting gas. No secondary discharges are observed while operating the detector in this configuration, even if $E_{\textrm{ind}}$ is set to a value at which secondary discharges are observed frequently in the standard configuration (\figrefbra{sec:results:fig:GDTsetups:a}). This demonstrates that the secondary discharge is not a response of the biasing circuit to a discharge occurring in the circuit, but it is caused by an effect in the gas. 

\subsection{Secondary discharges in the transfer gap}
\label{sec:results:subsec:2ndarysInTGap}
\subsubsection{General observations}
\begin{figure}
\centering
\includegraphics[width=0.49\textwidth, bb = 0 0 720 538, clip = true]{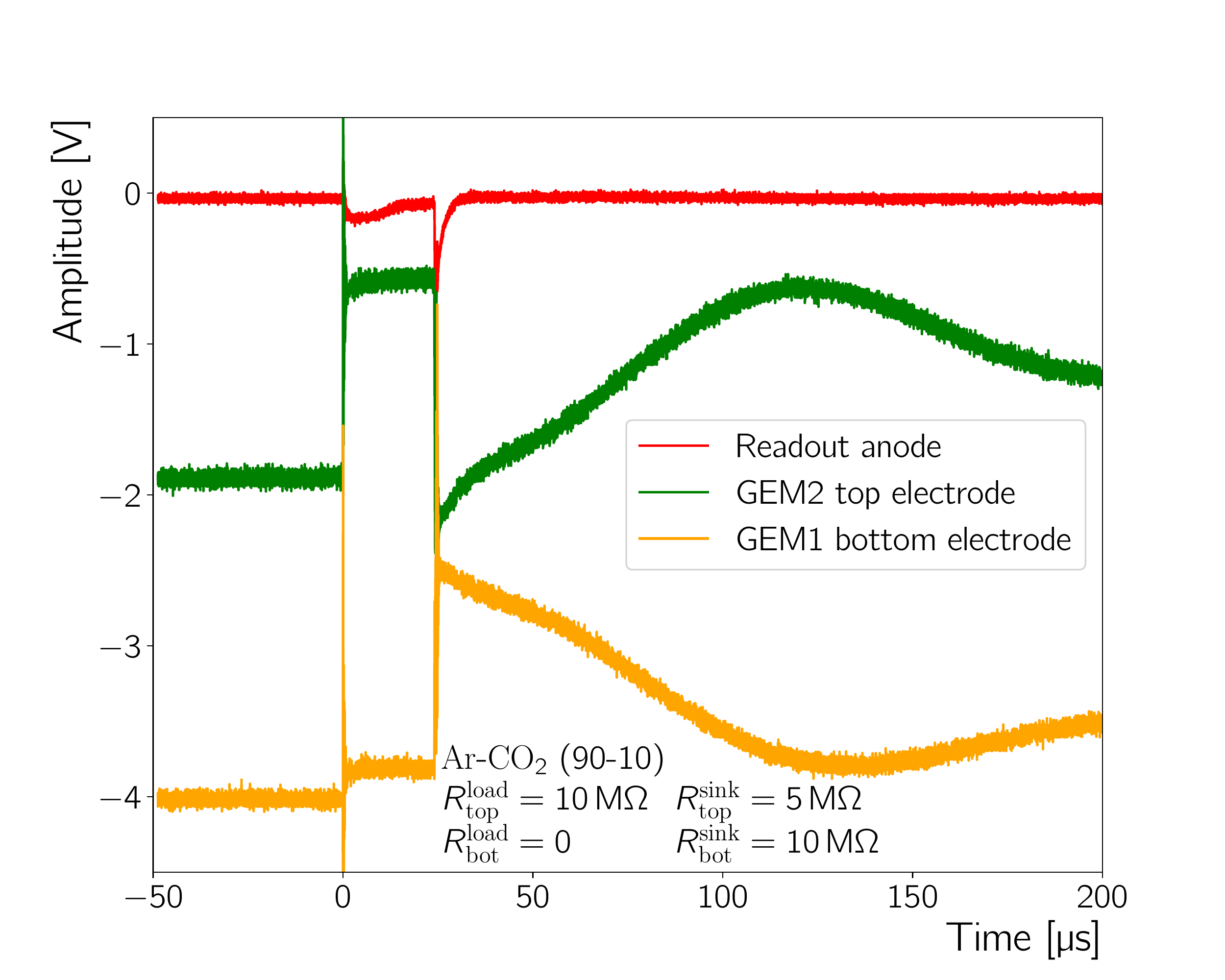}
\caption{\label{sec:results:fig:HVprobeSignals:readoutAndGEMPotentialsSecDc:etgem2tgem1b}(Colour online) Measurement of the potential during a primary discharge in GEM2 and secondary discharges in the gap between GEM1 and GEM2. The difference between the high voltage probe signals measured at the GEM2 top electrode and at the GEM1 bottom electrode illustrates the change of the transfer field during the primary and secondary discharge.}
\end{figure}
The set-up depicted in \figref{sec:setup:fig:setup} and described in \secref{sec:setup} is extended by another GEM foil, thus allowing to study (secondary) discharges in a double GEM set-up. In the double-GEM case we refer to the GEM closer to the cathode as GEM1 and to the GEM closer to the readout anode as GEM2. Primary discharges are induced in GEM2 by using appropriate HV settings (see \secrefbra{sec:setup}). These primary discharges change the transfer field between GEM2 and GEM1, because of the drop of the GEM2 top potential, which is approximately $\Delta V_{\textrm{GEM2}}$ when $R^{\textrm{load}}_{\textrm{bot}}=0$. \Figref{sec:results:fig:HVprobeSignals:readoutAndGEMPotentialsSecDc:etgem2tgem1b} illustrates this between 0 and $\sim\,\SI{20}{\micro\second}$. The GEM1 bottom potential drops by a lesser voltage due to capacitive coupling of this GEM electrode to the GEM2 top electrode (Figs.~\ref{sec:results:fig:HVprobeSignals:readoutAndGEMPotentialsSecDc:etgem2tgem1b} and \ref{sec:results:fig:HVprobeSignals:readoutAndGEMPotentialsSecDc:dcAndNoDcInGEM1}). When the secondary discharge occurs ($t\sim\,\SI{20}{\micro\second}$ in \figrefbra{sec:results:fig:HVprobeSignals:readoutAndGEMPotentialsSecDc:etgem2tgem1b}), the GEM2 top and the GEM1 bottom potentials approach each other in less than \SI{1}{\micro\second} until they reach approximately the same voltage. This is analogue to the observation that the GEM2 bottom potential approaches the ground potential during a secondary discharge in the induction gap (see \secrefbra{sec:results:subsec:HVProbeSignals}) and it indicates that the secondary discharge is the breakdown of the concerned gap. No field change or precursor is observed before this breakdown of the gap.

\subsubsection{Secondary discharges in the transfer gap: Caused by GEM1 or GEM2?}
\label{sec:results:subsec:sourceOfGEM1discharges}
Each secondary discharge in the induction gap is triggered by a discharge of the GEM facing the readout anode. In case of a multi-GEM system, the phenomenon of discharge propagation from one GEM to another has to be considered. A discharge in one GEM can be propagated to another GEM by \textit{e.g.} the photons created during the primary discharge which reach the other GEM and can create ionisations \cite{Deisting:2018gjp}. In this section we will discuss this phenomenon and its relevance for secondary discharges qualitatively. In \secref{sec:results:subsec:onsetInTransferGap} the arguments made here are quantified, considering the discharge propagation probability $P_{\textrm{prop}}$ describing the fraction of the primary GEM2 discharges which are propagated to GEM1.\\ \indent
\begin{figure}
\centering
\subfloat[]{\includegraphics[width=0.49\textwidth, bb = 0 0 720 538, clip = true]{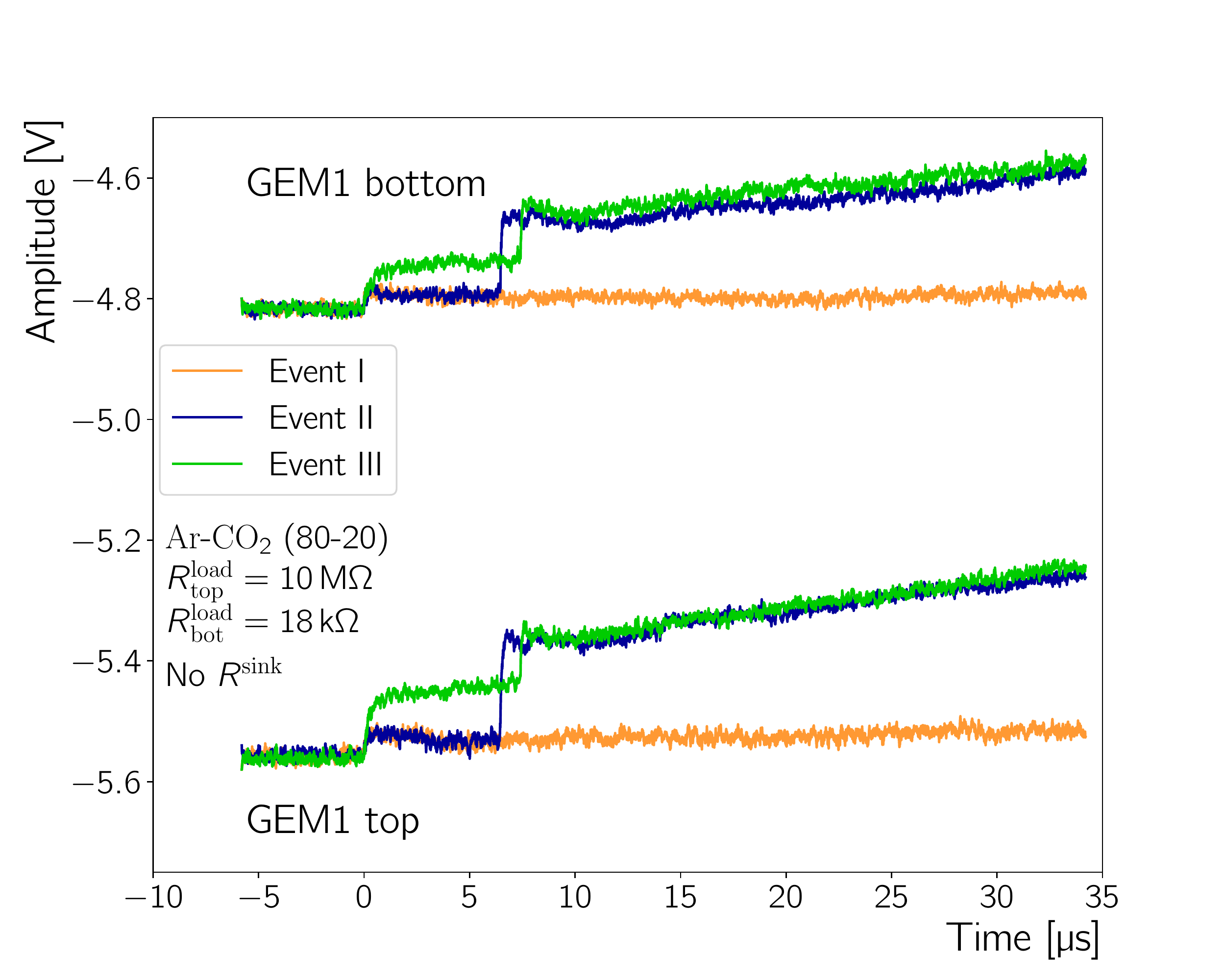}\label{sec:results:fig:HVprobeSignals:readoutAndGEMPotentialsSecDc:etgem1tgem1b}}\\
\subfloat[]{\includegraphics[width=0.40\textwidth, bb = 0 0 658 497, clip = true]{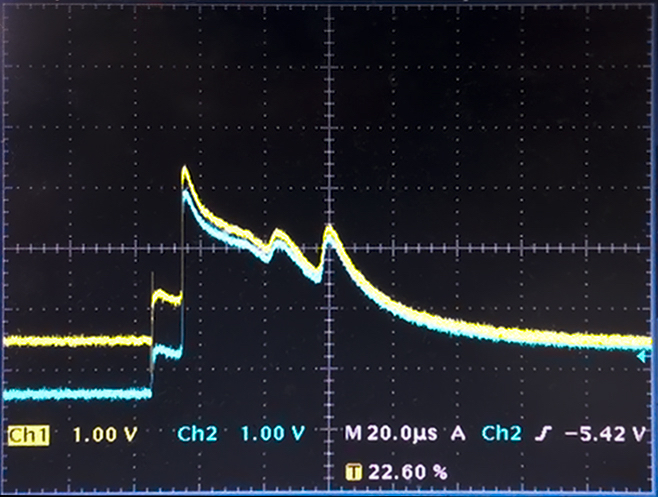}\label{sec:results:fig:HVprobeSignals:oscillogrammGEM1TopBottomRDec100k}}
\caption{\label{sec:results:fig:HVprobeSignals:readoutAndGEMPotentialsSecDc:dcAndNoDcInGEM1}(Colour online) Measurements of the GEM1 potentials during a primary discharge in GEM2 with and without a subsequent secondary discharge in the transfer gap between the GEMs. \protect\subref{sec:results:fig:HVprobeSignals:readoutAndGEMPotentialsSecDc:etgem1tgem1b} Three typical events without (I and II) and with (III) a discharge in GEM1, \textit{cf.} \secref{sec:results:subsec:sourceOfGEM1discharges} for an explanation. \protect\subref{sec:results:fig:HVprobeSignals:oscillogrammGEM1TopBottomRDec100k} Potential measurement at the bottom (Ch1) and top (Ch2) electrode of GEM1. During the discharge $\Delta V_{\textrm{GEM1}}$ does not change, however, at the time of the secondary discharge GEM1 discharges as well (\arcois{90-10}; $R_{\textrm{top}}^{\textrm{load}}=\SI{10}{\mega\ohm}$; $R^{\textrm{sink}}_{\textrm{top}}=\SI{5}{\mega\ohm}$; $R^{\textrm{sink}}_{\textrm{bot}}=\SI{10}{\mega\ohm}$; $R_{\textrm{bot}}^{\textrm{load}}=\SI{100}{\kilo\ohm}$).}
\end{figure}
\begin{figure*}
\centering
\subfloat[]{\includegraphics[width=0.49\textwidth, bb = 0 0 720 538, clip = true]{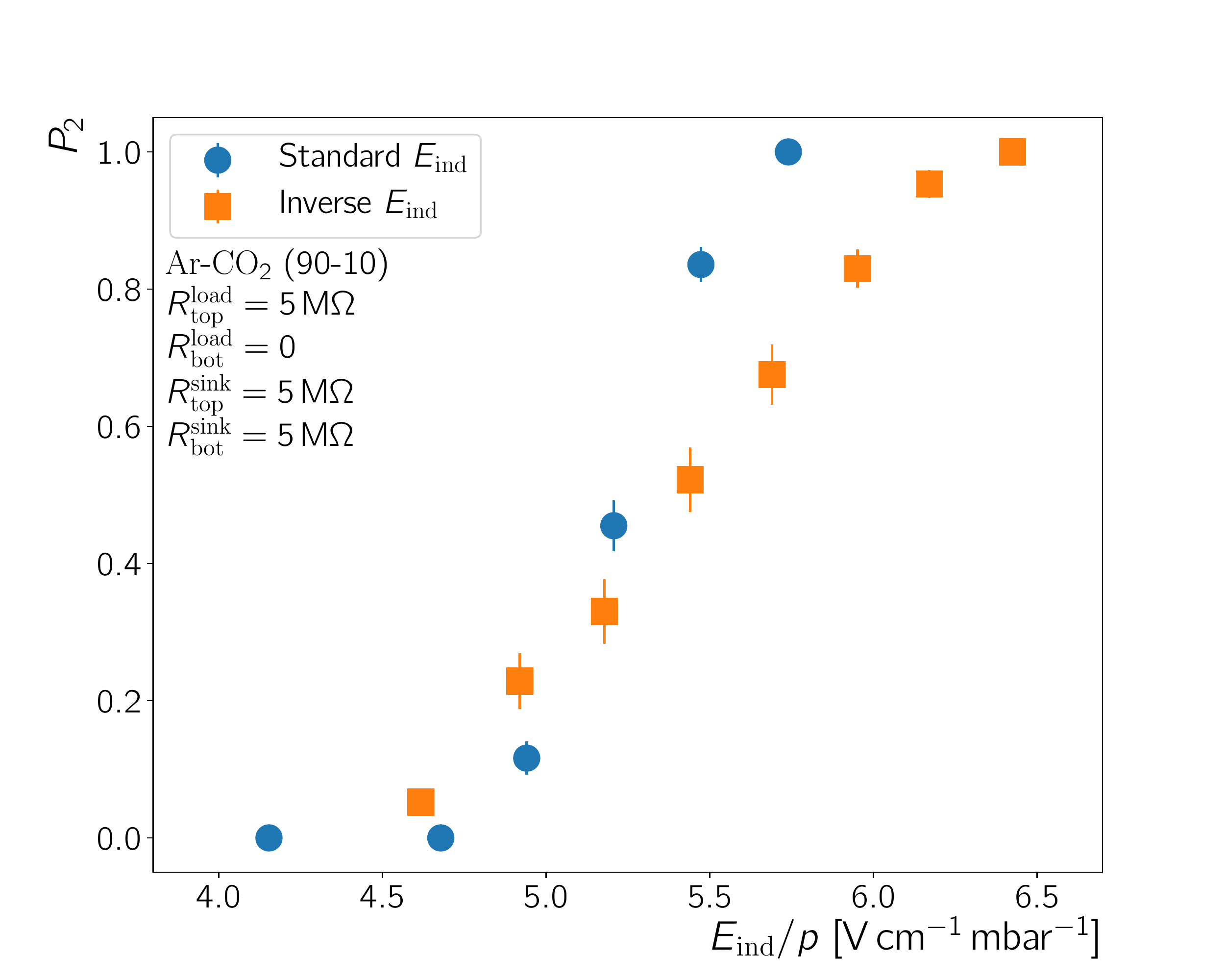}\label{sec:results:fig:norm_vs_revers_onset}}
\subfloat[]{\includegraphics[width=0.49\textwidth, bb = 0 0 720 538, clip = true]{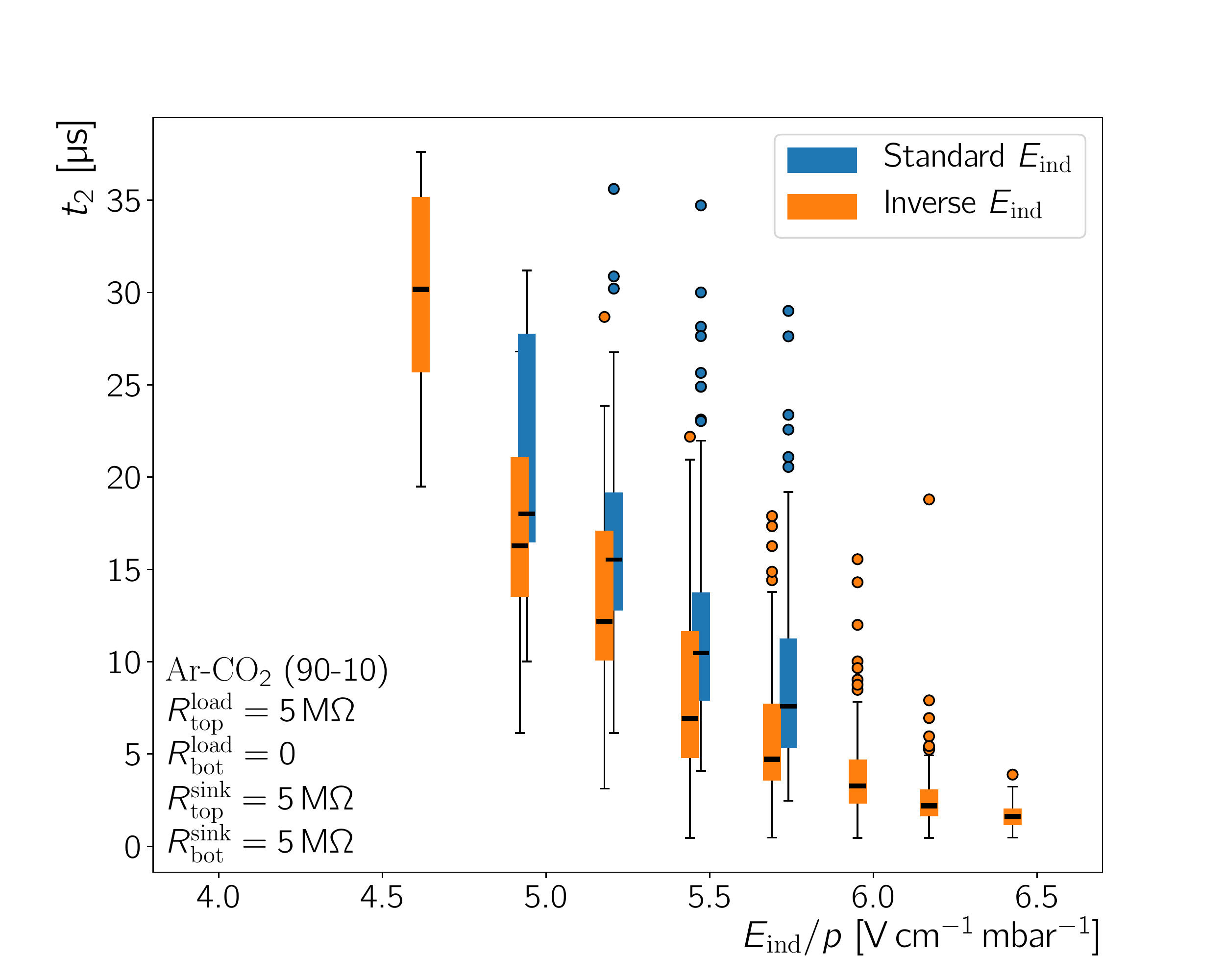}\label{sec:results:fig:norm_vs_revers_times}}
\caption{\label{sec:results:fig:norm_vs_revers_onset_and_times}(Colour online) \protect\subref{sec:results:fig:norm_vs_revers_onset} The probability to observe a secondary discharge after a primary discharge as function of the induction field for normal and inverted fields. \protect\subref{sec:results:fig:norm_vs_revers_times} For the same fields, time difference between primary and secondary discharges ($t_2$). The median of the $t_2$ values for a certain field is indicated by the horizontal bar on the box. For further explanations see the text.}
\end{figure*}
Depending on whether a secondary discharge in the transfer gap is triggered by a discharge in GEM2 or GEM1, the secondary discharge evolves against or with the transfer field direction. \Figref{sec:results:fig:HVprobeSignals:readoutAndGEMPotentialsSecDc:etgem1tgem1b} shows three different cases during which the primary discharges take place in GEM2:
\begin{enumerate}[I]
	\item $\ \ $The primary discharge in GEM2 is not propagated to GEM1 and no secondary discharge occurs in the transfer gap.
	\item $\ \ $The primary discharge in GEM2 is not propagated to GEM1 and a secondary discharge occurs in the transfer gap.
	\item $\ \ $The primary discharge in GEM2 is propagated to GEM1 and a secondary discharge occurs in the transfer gap.
\end{enumerate} 
Note that the particular HV probes used in \figref{sec:results:fig:HVprobeSignals:readoutAndGEMPotentialsSecDc:etgem1tgem1b} do not follow fast voltage drops accurately as occurring during a (secondary) discharge, hence, they do not allow to make measurements of the real DC voltage during such events. However, they allow to make relative comparisons and qualitative statements, \textit{e.g.} whether there was a voltage drop or not.\footnote{The quality of the probes' DC voltage measurement during fast voltage changes has been assessed with known signals and with tests of different HV-probes in otherwise similar discharge settings.}\\ \indent
For I and II a similar voltage drop on the top and bottom GEM1 electrode is observed. These simultaneous drops are caused by the capacitive coupling between GEM1 and GEM2 and the capacitative coupling between the GEM1 electrodes, similar to what has been discussed for \figref{sec:results:fig:HVprobeSignals:readoutAndGEMPotentialsSecDc:etgem2tgem1b}. In case III the drop of the GEM1 top electrode potential is larger than the drop of the GEM1 bottom electrode potential, indicating that the GEM has discharged and $\Delta V_{\textrm{GEM1}}$ is reduced. A secondary discharge is observed around $t\sim\,\SI{7}{\micro\second}$ in the events exemplifying case II and III.\footnote{The voltage drop measured during the secondary discharges depicted in \figref{sec:results:fig:HVprobeSignals:readoutAndGEMPotentialsSecDc:etgem1tgem1b} is lower than the corresponding drop shown in Figures~\ref{sec:results:fig:HVprobeSignals:readoutAndGEMPotentialsSecDc:etgem2tgem1b} and \ref{sec:results:fig:HVprobeSignals:oscillogrammGEM1TopBottomRDec100k}. This is an artefact of the different HV probes used for the measurements in \figref{sec:results:fig:HVprobeSignals:readoutAndGEMPotentialsSecDc:etgem1tgem1b}.} GEM1 did not discharge in case II, indicating that the secondary discharge in the transfer gap was initiated by a primary discharge in the GEM below (GEM2). Similarly, \figref{sec:results:fig:HVprobeSignals:oscillogrammGEM1TopBottomRDec100k} shows another event where GEM1 does not discharge in advance of a secondary discharge in the gap below. This shows that the secondary discharges do not only evolve in the gap below a GEM (\textit{i.e.} case III and \textit{cf.} \secs\ref{sec:results:subsec:primaryAndSecondary}--\ref{sec:results:subsec:HVProbeSignals}), but they can also evolve against the field direction through a transfer gap (\textit{i.e.} II). Hence, both GEMs facing the transfer gap can be responsible for secondary discharges.

%% file: onsetAndTimeCurves.tex
\section{Occurrence probability of secondary discharges}
\label{sec:results:onsetcurves}
The probability $P_2$ to observe a secondary after a primary discharge is defined as $P_2 = \frac{N_2}{N_1}$, where $N_1$ and $N_2$ is the number of primary and secondary discharges, respectively. In order to perform a systematic measurement of $P_2$, certain HV settings are applied to the detector and for a given time primary and secondary discharges are counted using the logic described in \secref{sec:setup}. Then the induction field (or transfer field in case of double GEM measurements) is changed and the next measurement is performed.
\begin{figure*}
\centering
\subfloat[]{\includegraphics[width=0.49\textwidth, bb = 0 0 720 538, clip = true]{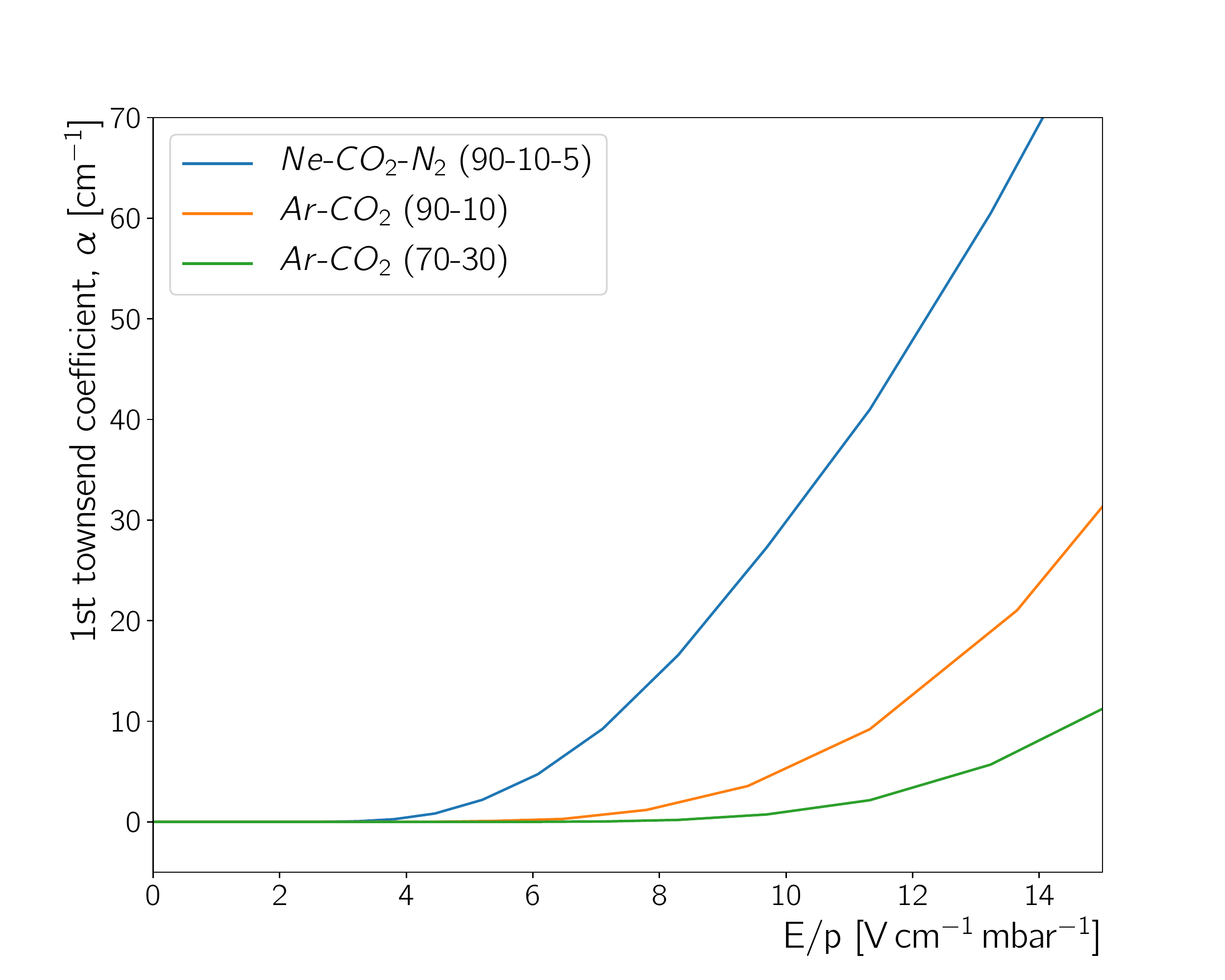}\label{sec:results:fig:townsend}}
\subfloat[]{\includegraphics[width=0.49\textwidth, bb = 0 0 720 538, clip = true]{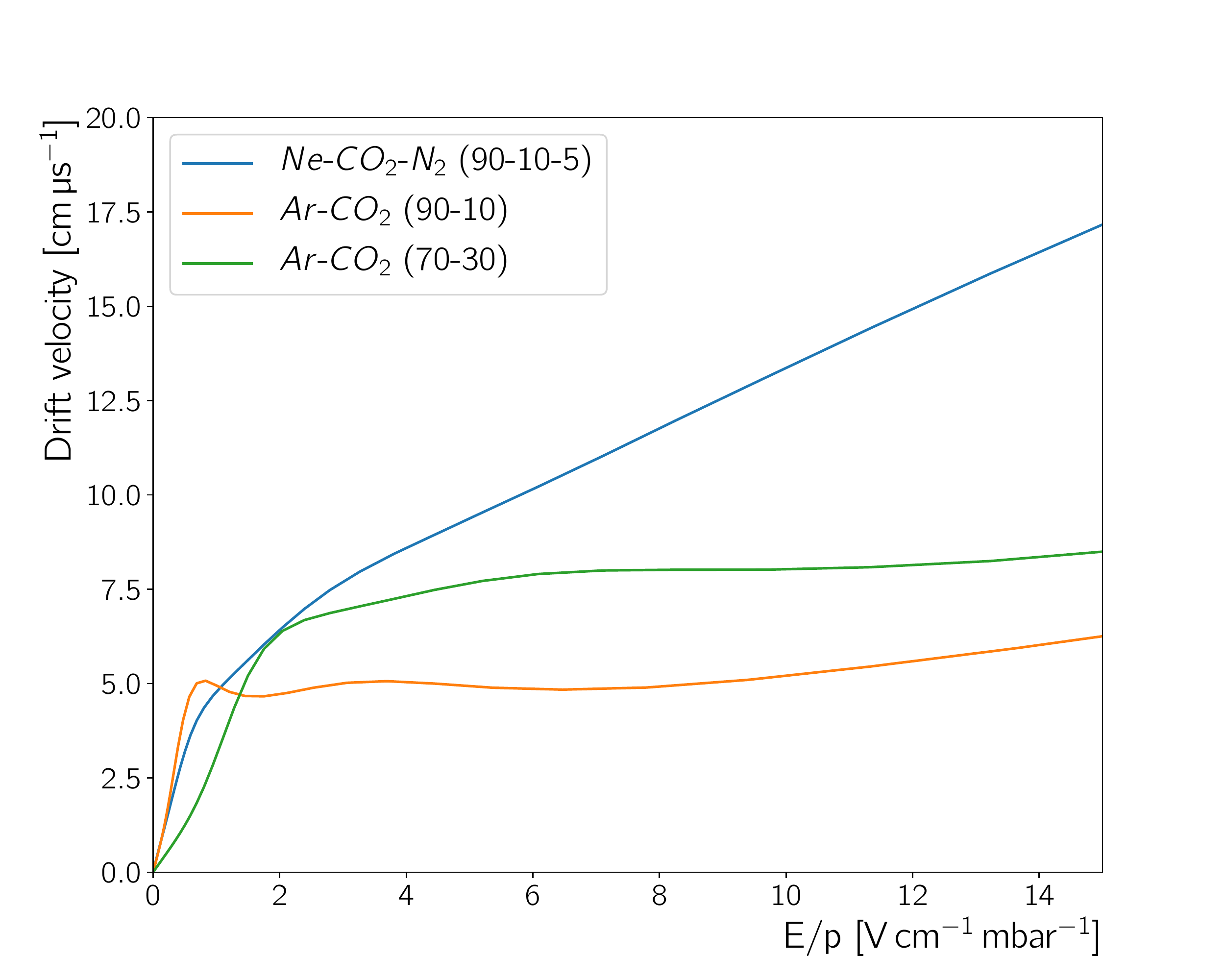}\label{sec:results:e-vdrift}}
\caption{\label{sec:results:fig:townsendAnde-vdrift}(Colour online) \textsc{Magboltz} \cite{Biagi1018382} simulation \protect\subref{sec:results:fig:townsend} of the first Townsend coefficient and \protect\subref{sec:results:e-vdrift} the electron drift velocity. The simulation has been done for a gas pressure of \SI{1}{\bar} and a temperature of \SI{300}{\kelvin}.}
\end{figure*}

\subsection{Onset of secondary discharges in the induction gap}
\label{sec:results:onsetcurves:indgap}
Such a series of measurements in an \arcois{90-10} mixture is displayed in \figref{sec:results:fig:norm_vs_revers_onset}, showing that $P_2$  depends on the value of the induction field. Binomial errors are used here and hence forth for the uncertainty on the probability.\\ \indent
As can be seen in \figref{sec:results:fig:norm_vs_revers_onset}, up to a certain threshold of $E_{\textrm{ind}}$, no secondary discharge takes place or the number of secondary discharges is negligible. By increasing the field above this threshold, secondary discharges occur, following the primary ones. At higher values of the induction field, $P_2$ reaches eventually unity, where every primary discharge is followed by the breakdown of the induction gap. This confirms previous results, showing that these discharges only occur if the electric field below a discharging GEM is high enough \cite{Bachmann}.\\ \indent
The electric field at which secondary discharges occur is lower than the electric field necessary for gas amplification. This is particularly interesting to note, since a breakdown occurs in the respective gap during a secondary discharge, see \secref{sec:results:subsec:HVProbeSignals}. The first Townsend coefficient becomes larger than zero at fields substantially higher than those necessary to trigger secondary discharges (\figrefbra{sec:results:fig:townsend}). A mechanism different from the Townsend discharge is therefore needed to explain the occurrence of secondary discharges.

\subsubsection{Reverse field direction}
\label{sec:results:invertedEInd}
Secondary discharges in the transfer gap of a detector can be caused by either GEM facing it, as pointed out in \secref{sec:results:subsec:sourceOfGEM1discharges}, showing that secondary discharges can evolve along and against the field direction. We observe secondary discharges in the induction gap as well, when the direction of the induction field is reversed. $P_2$ as a function of the absolute value of the inverted $E_{\textrm{ind}}$ is similar to the case of the normal field direction, as can be seen in \figref{sec:results:fig:norm_vs_revers_onset}. This indicates that the propagation depends to first order only on the strength of the electric field below a GEM and not on the field direction. 

\subsubsection{Time between primary and secondary discharge}
\label{sec:results:timeBetweenPrimaryAndSecondary}
An interesting characteristic of the secondary discharge phenomena is the relatively long time between a primary discharge and the occurrence of the secondary one ($t_2$) of up to several \SI{10}{\micro\second} (\textit{e.g.} Figures~\ref{sec:results:fig:HVprobeSignals:GDTs:config1HV} and \ref{sec:results:fig:HVprobeSignals:readoutAndGEMPotentialsSecDc:etgem2tgem1b}). We investigate this by using waveforms recorded by the oscilloscope to measure and compare $t_2$ for the case of the normal and inverted field direction. For different HV settings the corresponding waveforms have been analysed and the time between primary and secondary discharge has been extracted for each such event. The resulting $t_2$ distributions for each HV setting feature a clear peak with a tail towards large values. For this reason we display the measured times for a single $E_{\textrm{ind}}$ as box plot (\figrefbra{sec:results:fig:norm_vs_revers_times}). The box marks the $t_{2}$ values in the region form the \SI{25}{\%} to \SI{75}{\%} percentile, while the median $t_{2}$ is marked by a horizontal bar on that box. Furthermore the whiskers show the region corresponding to the \SI{25}{\%} (\SI{75}{\%}, respectively) percentile to \SI{25}{\%} minus (\SI{75}{\%} plus, respectively) the difference between the \SI{75}{\%} and \SI{25}{\%} percentile $t_{2}$. All measured $t_2$ value outside the range of the whiskers are indicated as dots. \Figref{sec:results:fig:norm_vs_revers_times} shows that the median $t_2$ decreases in a similar manner for the case of the normal and inverted $E_{\textrm{ind}}$ direction with increasing modulus of the induction field. In addition the full $t_2$ distribution shifts towards lower values at the same time. The median $t_2$ values measured for the inverted field direction are systematically higher than the ones for the standard field direction, however this difference is not significant.\\ \indent
The electron drift time across the induction gap (\SI{2}{\milli\meter}) can be calculated using electron drift velocities in \figref{sec:results:e-vdrift}. At the fields and gas conditions employed for this work the time is about $\SI{0.05}{\micro\second}$. The median time between primary and secondary discharge is long compared to that, but it is of the same order of magnitude as the time ions would need to cross a \SI{2}{\milli\meter} gap. \textit{E.g.} for our gas conditions and typical fields at which secondary discharges are observed, we calculate a time of $\sim\,\SI{22}{\micro\second}$ using the ion mobilities in \cite{DEISTING20181}. However, the observed decrease of $t_2$ for a relatively small change of $E_{\textrm{ind}}$ is much larger than the expected corresponding change of the ion drift time. This, as well as the fact that secondary discharges are observed for both field directions (\secs\ref{sec:results:subsec:sourceOfGEM1discharges} and \ref{sec:results:invertedEInd}), suggests that the secondary discharges are not caused by charges created at the time of the primary discharge which then cross the concerned gap.

\subsubsection{Onset for different gas mixtures}
\label{sec:results:subsubsec:differentMixtures}
\begin{figure}
\centering
\subfloat[]{\includegraphics[width=0.49\textwidth, bb = 0 0 720 538, clip = true]{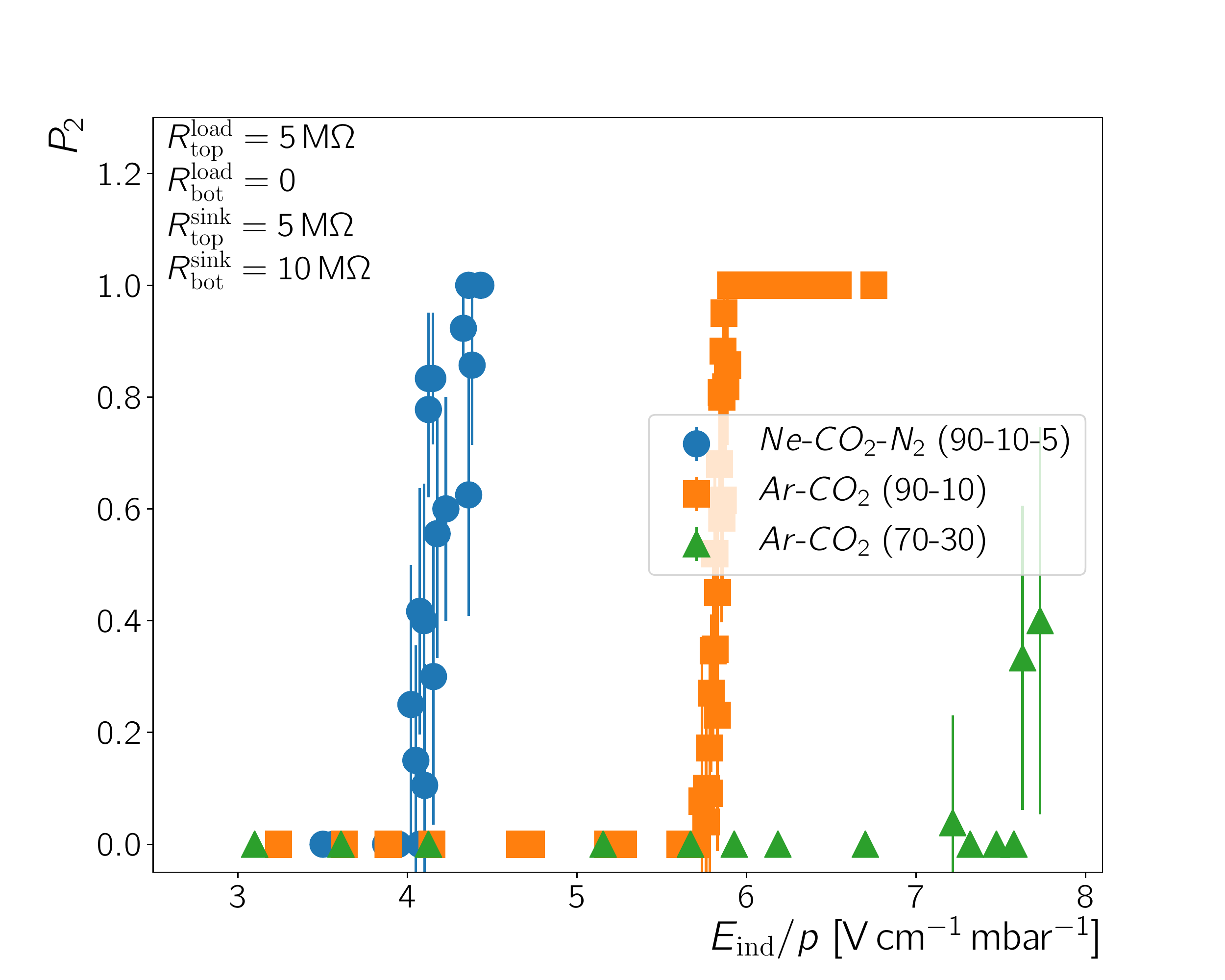}\label{sec:results:fig:onset_for_different_mixtures:piotr}}\\
\subfloat[]{\includegraphics[width=0.49\textwidth, bb = 0 0 720 538, clip = true]{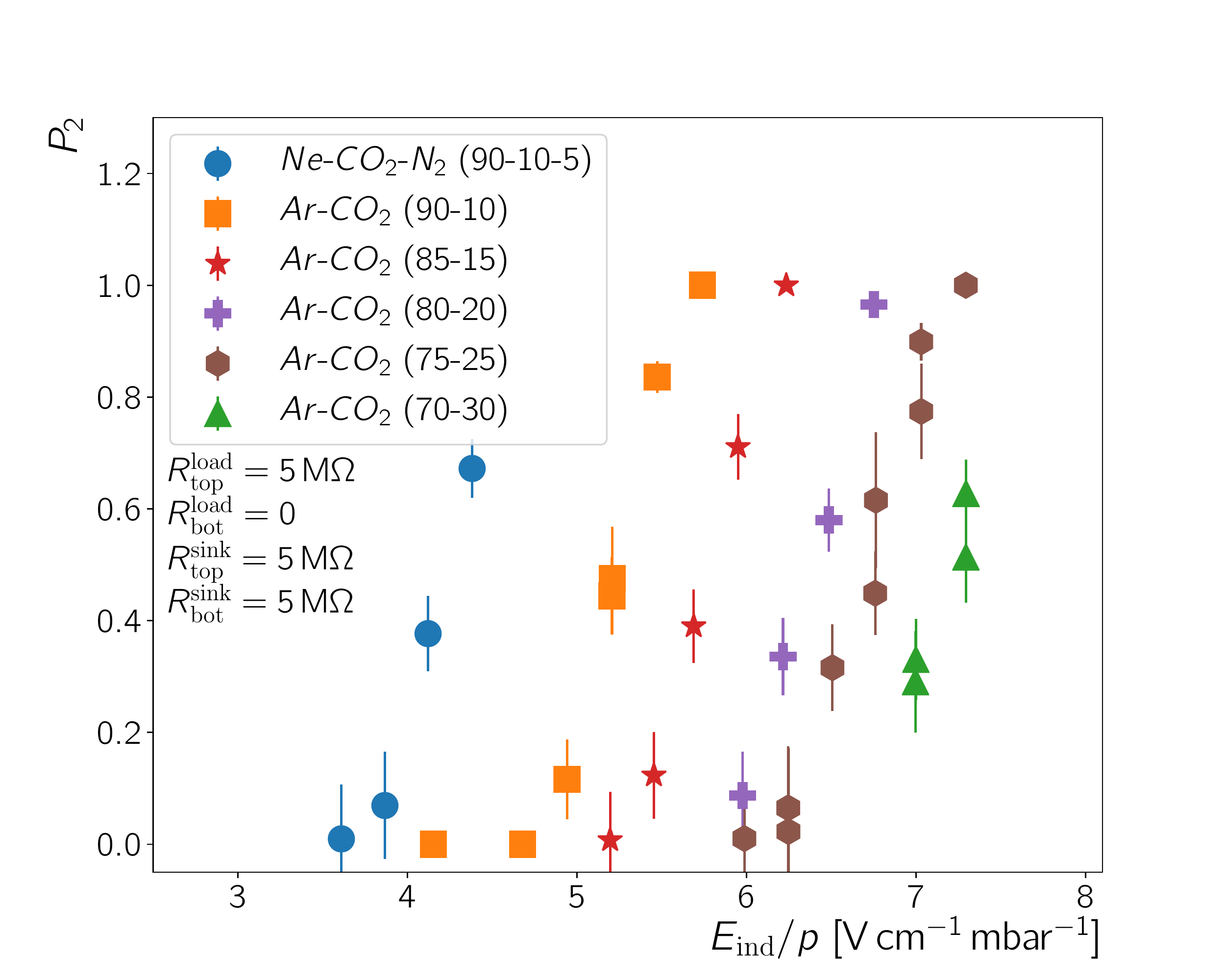}\label{sec:results:fig:onset_for_different_mixtures:daniel}}
\caption{\label{sec:results:fig:onset_for_different_mixtures}(Colour online) Onset of secondary discharges in the induction gap for different gas mixtures. All measurements within each plot are performed with the same GEM, however, the GEM used in plot \protect\subref{sec:results:fig:onset_for_different_mixtures:piotr} and in \protect\subref{sec:results:fig:onset_for_different_mixtures:daniel} differ. Furthermore, the GEM of plot \protect\subref{sec:results:fig:onset_for_different_mixtures:piotr} is different from the GEMs used for the other plots in this work. The hardware configuration -- except for the GEM -- is the same for all measurements.}
\end{figure}
\begin{table*}
\centering
\begin{tabular}{l|c|c}
	           & $E_{\textrm{ind}}^{\textrm{on}}$ & $E_{\textrm{ind}}^{\textrm{on}}$ \\
Gas mixture    & (\figrefbra{sec:results:fig:onset_for_different_mixtures:piotr}) & (\figrefbra{sec:results:fig:onset_for_different_mixtures:daniel}) \\ \hline
\baseline{}    & \SI{4.2(3)}{\volt\per\centi\meter\per\milli\bar} & \SI{4.3(2)}{\volt\per\centi\meter\per\milli\bar} \\
\arcois{90-10} & \SI{5.7(1)}{\volt\per\centi\meter\per\milli\bar} & \SI{5.2(1)}{\volt\per\centi\meter\per\milli\bar} \\
\arcois{85-15} & & \SI{5.8(2)}{\volt\per\centi\meter\per\milli\bar} \\
\arcois{80-20} & & \SI{6.3(2)}{\volt\per\centi\meter\per\milli\bar} \\
\arcois{75-25} & & \SI{6.7(3)}{\volt\per\centi\meter\per\milli\bar} \\
\arcois{70-30} & $>\SI{7.5}{\volt\per\centi\meter\per\milli\bar}$ & \SI{7.2(3)}{\volt\per\centi\meter\per\milli\bar} \\
\end{tabular}
\caption{\label{sec:results:tab:onset_for_different_mixtures}The onset field for secondary discharges in the induction gap, $E_{\textrm{ind}}^{\textrm{on}}$, (field at which $P_2=0.5$) derived from the measurements in \figref{sec:results:fig:onset_for_different_mixtures}. The difference in $E_{\textrm{ind}}^{\textrm{on}}$ for the same mixtures are due to the different GEMs used for the measurements shown in Figures~\ref{sec:results:fig:onset_for_different_mixtures:piotr} and \ref{sec:results:fig:onset_for_different_mixtures:daniel}.}
\end{table*}
We define the onset field, $E^{\textrm{on}}_{k}$ where $k\in (\textrm{ind}, \textrm{t})$, as the electric field at which $P_2=0.5$, \textit{i.e.} the probability for a secondary discharge to occur after a primary discharge is \SI{50}{\%}. Figures~\ref{sec:results:fig:onset_for_different_mixtures:piotr} and \ref{sec:results:fig:onset_for_different_mixtures:daniel} show that $E^{\textrm{on}}_{\textrm{ind}}$ changes for different gas mixtures, when the hardware configuration remains the same. The difference in $E^{\textrm{on}}_{\textrm{ind}}$ for the same gas mixture between the two plots (\figrefbra{sec:results:fig:onset_for_different_mixtures:piotr} and \ref{sec:results:fig:onset_for_different_mixtures:daniel}) is due to the different GEM used for the measurements. We excluded other effects as \textit{e.g.} differences in the induction gap length introduced during the GEM exchange and also in previous works it has been observed that the exact value of the onset field is different for individual GEMs \cite{Deisting:2018gjp,pgasik2016rd51,baDatz2017}. Table \ref{sec:results:tab:onset_for_different_mixtures} lists the onset fields measured for the different mixtures. Comparing these mixtures, the ordering of $E^{\textrm{on}}_{\textrm{ind}}$ with quencher concentration follows the order as expected from the mixtures' Townsend coefficients (\figrefbra{sec:results:fig:townsend}). However, for all mixtures $E^{\textrm{on}}_{\textrm{ind}}$ is lower than the field needed for effective gas amplification. A similar observation has been made for secondary discharges in the transfer gap \cite{Deisting:2018gjp}.

\subsubsection{Influence of the drift field}
\label{sec:results:subsubsec:influenceOfTheDriftField}
We excluded that the secondary discharge is due to ions crossing the induction (transfer) gap. Nevertheless, ions created during the primary discharge could be a possible cause of secondary discharges. If these ions drift, despite the absence of field across the GEM, through the hole, they would eventually impinge on the top copper layer. There, they can extract electrons from the top GEM electrode, which could seed the secondary discharge and could also explain the current observed after the primary discharge (\secrefbra{sec:results:subsec:currentsBefore2NDary}).\\ \indent
\begin{figure*}
\centering
\subfloat[]{\includegraphics[width=0.49\textwidth, bb = 0 0 720 538, clip = true]{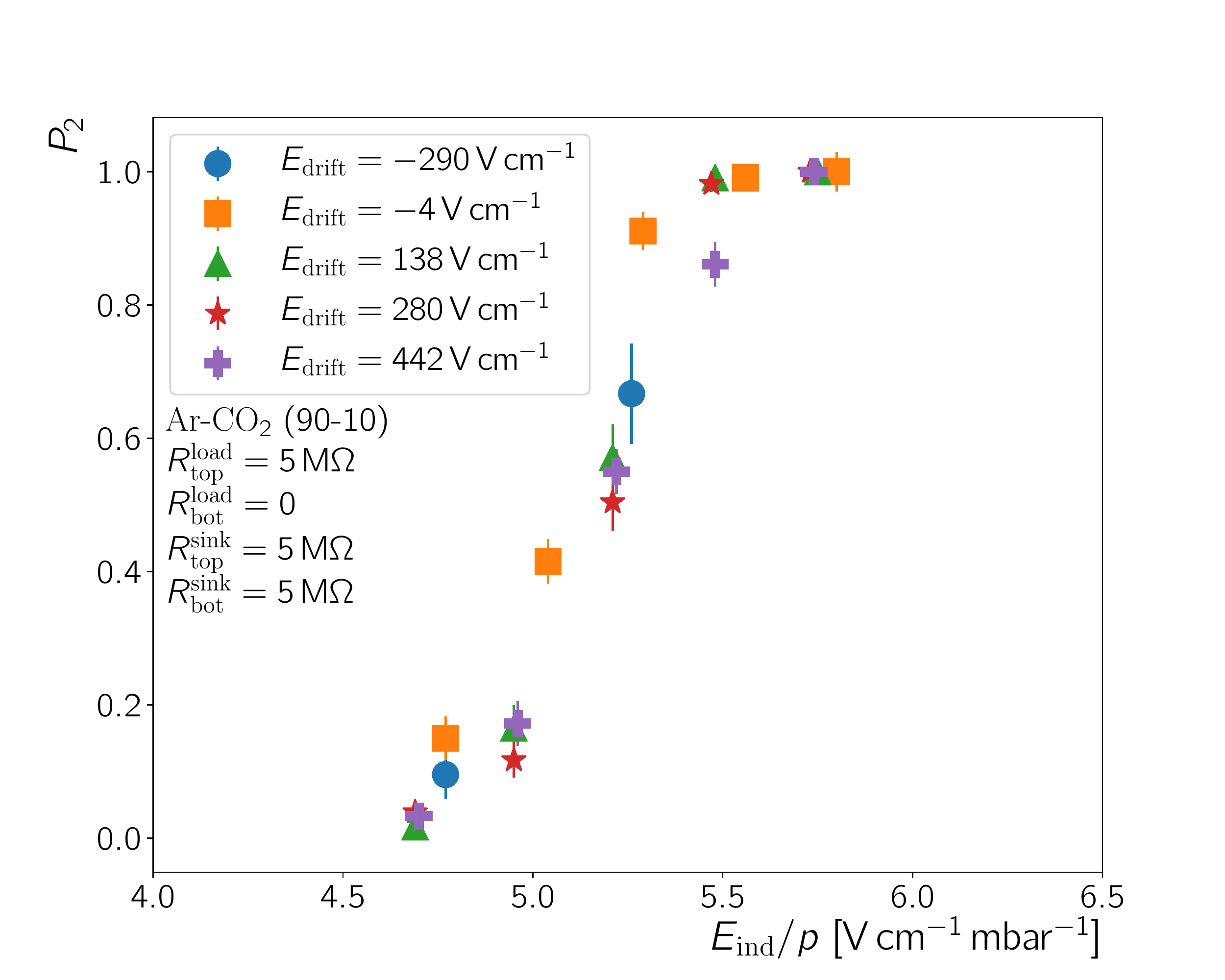}
\label{sec:results:fig:onset_for_different_drift}}
\subfloat[]{\includegraphics[width=0.49\textwidth, bb = 0 0 720 538, clip = true]{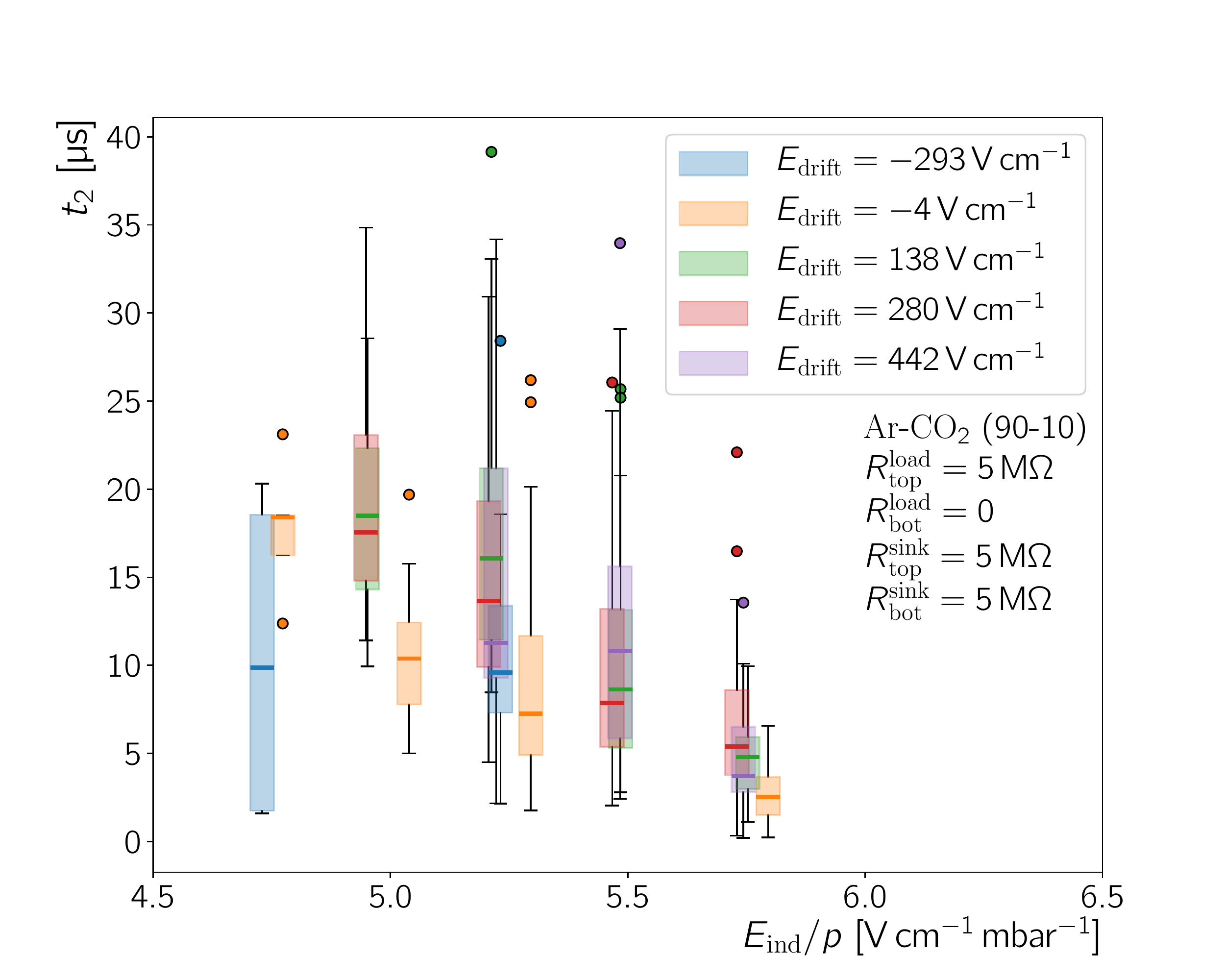}
\label{sec:results:fig:time_for_different_drift}}
\caption{\label{sec:results:fig:onset_and_time_for_different_drift}(Colour online) \protect\subref{sec:results:fig:onset_for_different_drift} $P_2$ as a function of $E_{\textrm{ind}}$ for different values of $E_{\textrm{drift}}$. The time difference ($t_2$) between primary and secondary discharges extracted from the waveforms recorded during the $P_2$ measurement is shown in \protect\subref{sec:results:fig:time_for_different_drift}. See the discussion of \figref{sec:results:fig:norm_vs_revers_times} for further explanations.}
\end{figure*}
To test this hypothesis we perform a measurement of $P_2$ varying the drift field above the GEM foil. The ion extraction efficiency from a GEM hole is expected to change over the range of fields employed for the measurements in \figref{sec:results:fig:onset_and_time_for_different_drift} \cite{killenberg2003modelling}. 
We observe no effect of a drift field variation in the range from $-\SI{290}{\volt\per\centi\meter}$ to $\SI{440}{\volt\per\centi\meter}$ on the  probability to observe a secondary discharge after a primary discharge as well as on the time differences between primary and secondary discharge. Thus we conclude that the ion extraction from a GEM hole does not play a role in the creation of the secondary discharge and it is therefore unlikely that secondary discharges are caused by ion bombardment of the top GEM electrode. 

\subsection{Mitigation of the secondary discharge onset using decoupling resistors}
\label{sec:results:mitigationWithDecouplingResistors}
\begin{figure*}
\centering
\subfloat[]{\includegraphics[width=0.49\textwidth, bb = 0 0 720 538, clip = true]{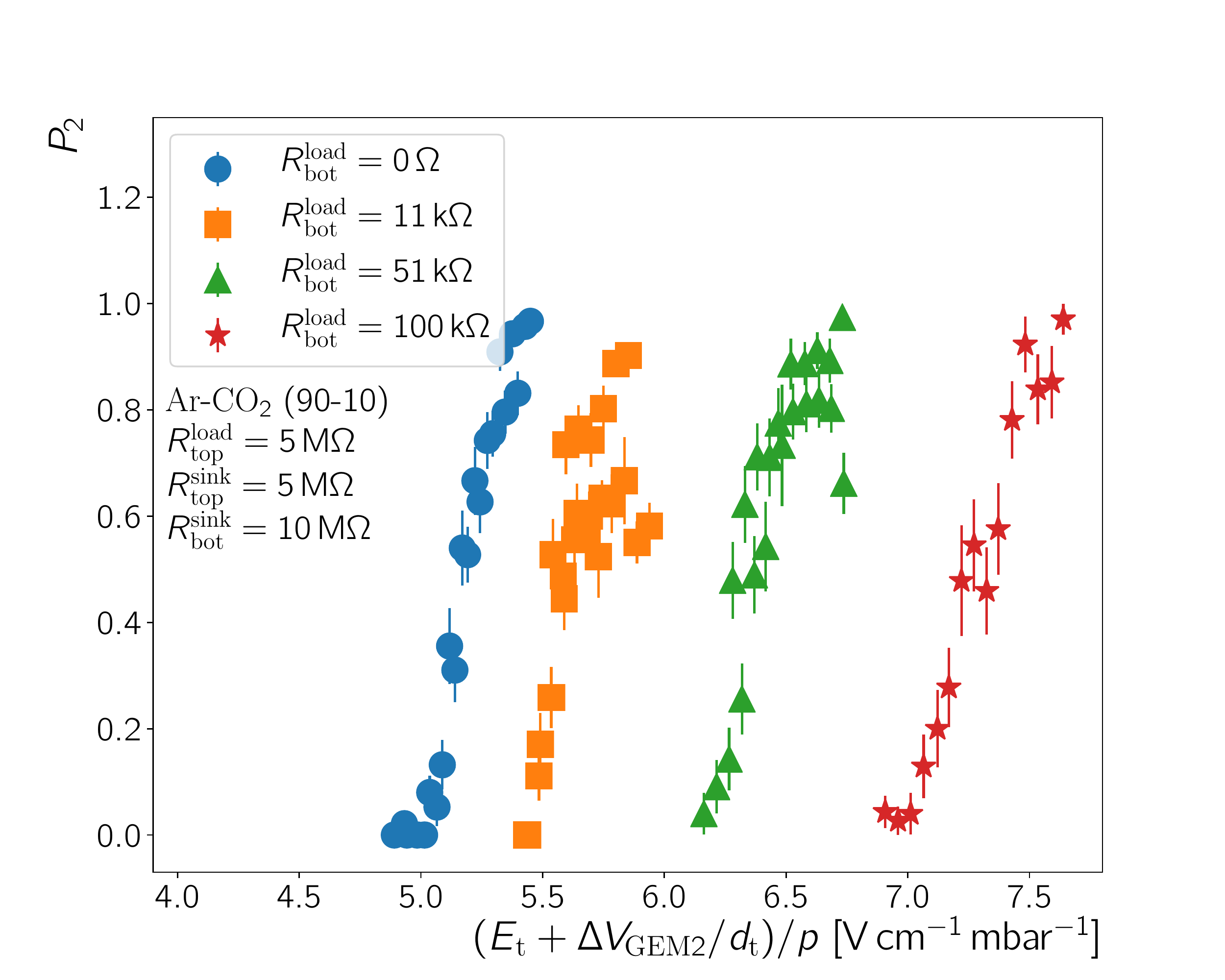}
\label{sec:results:fig:rBotLoadVsET}}
\subfloat[]{\includegraphics[width=0.49\textwidth, bb = 0 0 720 538, clip = true]{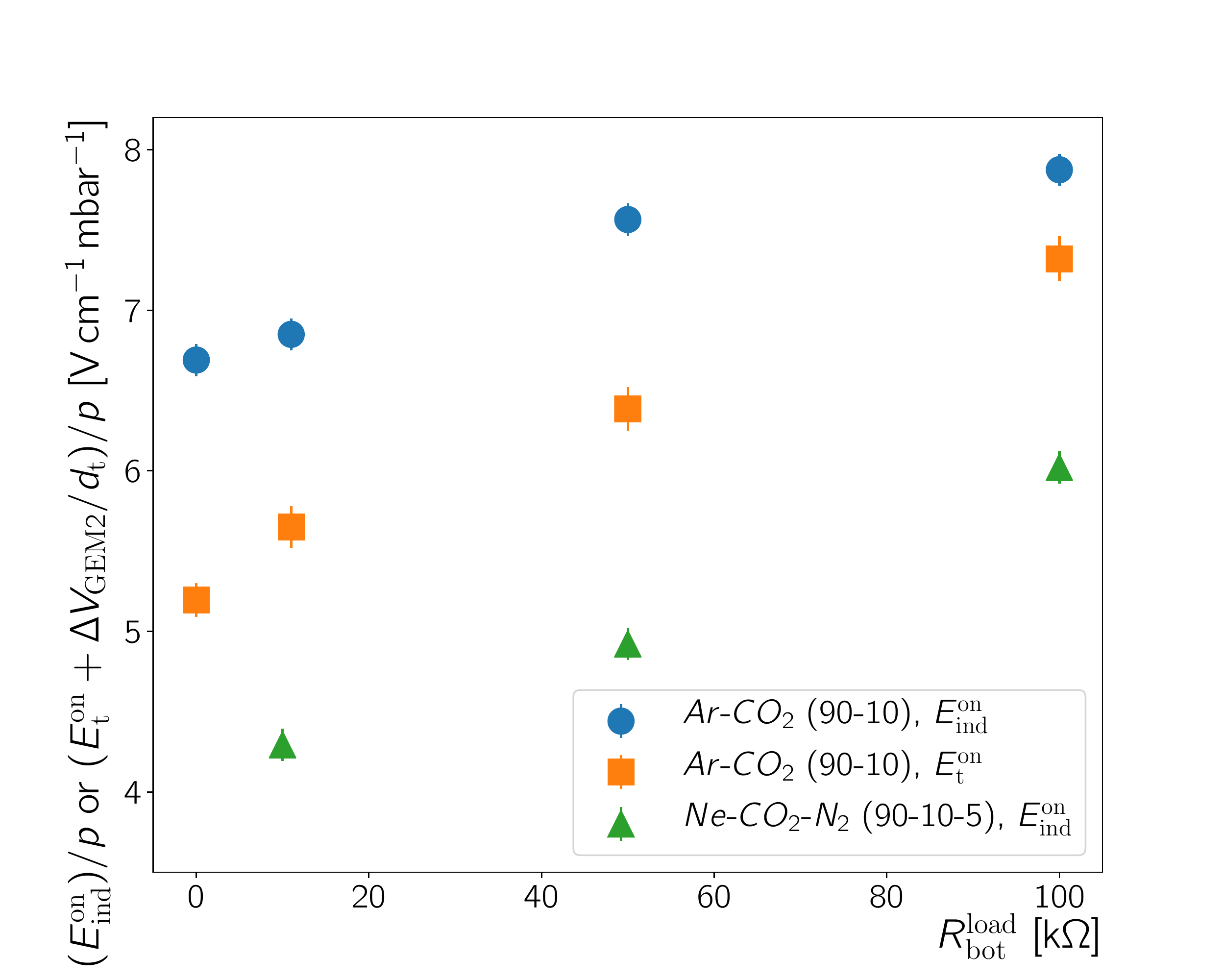}
\label{sec:results:fig:rBotLoadVsEOnsetIndTransArCO2andBaseline}}
\caption{\label{sec:results:fig:rBotLoadVsETAndEInd}(Colour online) \protect\subref{sec:results:fig:rBotLoadVsET} $P_2$ as a function of the sum of transfer field and the voltage across GEM2 divided by the transfer gap's length, $d_{\textrm{t}}$, ($\Delta V_{\textrm{GEM}}/d_{\textrm{t}}$) for different decoupling resistors $R_{\textrm{bot}}^{\textrm{load}}$. \protect\subref{sec:results:fig:rBotLoadVsEOnsetIndTransArCO2andBaseline} Onset field versus $R_{\textrm{bot}}^{\textrm{load}}$. In all cases the primary discharge is induced in the GEM close to the readout anode and the propagation probability is $P_{\textrm{prop}}\sim\,1$. The \baseline{} points are not recorded with the same GEMs as the other points, but with ones present during the \figref{sec:results:fig:onset_for_different_mixtures:piotr} measurements.}
\end{figure*}
An increase of the onset field for secondary discharges is found when a decoupling resistor is added to the HV supply path to the GEM bottom electrode, \textrm{i.e.} $R_{\textrm{bot}}^{\textrm{load}}\neq0$. \Figref{sec:results:fig:rBotLoadVsETAndEInd} shows several measurements of $P_2$ for different $R_{\textrm{bot}}$, while keeping the set-up otherwise identical. In \figref{sec:results:fig:rBotLoadVsEOnsetIndTransArCO2andBaseline}, the onset field, $E^{\textrm{on}}_{\textrm{ind}}$ and $E^{\textrm{on}}_{\textrm{t}}$, versus $R_{\textrm{bot}}^{\textrm{load}}$ is plotted. The $E^{\textrm{on}}_{\textrm{t}}$ in the figure refers to $E^{\textrm{on}}_{\textrm{t}}+\Delta V_{\textrm{GEM2}}/d_{\textrm{t}}$ and not just to the transfer field applied to the detector (see \secrefbra{sec:results:subsec:onsetInTransferGap}). These measurements show that $E^{\textrm{on}}$ for secondary discharges in the induction and in the transfer gap increases with increasing $R_{\textrm{bot}}$.\footnote{As mentioned in \secref{sec:results:subsec:sourceOfGEM1discharges} the discharge propagation from GEM2 to GEM1 has to be considered for secondary discharges in the transfer gap. For the measurements shown in \figref{sec:results:fig:rBotLoadVsETAndEInd} the voltage settings have been chosen such, that \SI{100}{\%} of the discharges in GEM2 are propagated to GEM1, \textit{i.e.} $P_{\textrm{prop}}=1$.} This effect is present in \arcois{90-10} and \baseline{}. For the latter, secondary discharges occur at fields at which one might desire to operate a detector (see \figrefbra{sec:results:fig:onset_for_different_mixtures}). For example the ALICE TPC GEM stacks will be operated in \baseline{} and the fields foreseen are as high as \SI{4}{\kilo\volt\per\centi\meter} \cite{aliceTpcUpgradeTDR2014}. Decoupling resistors are thus a suitable tool to make the occurrence of secondary discharges less likely.

\subsubsection{HV probe measurements while $R_{\textrm{bot}}\neq0$}
\label{sec:results:onsetcurves:subsec:decouplingR}
\begin{figure}
\centering
\includegraphics[width=0.49\textwidth, bb = 0 0 720 538, clip = true]{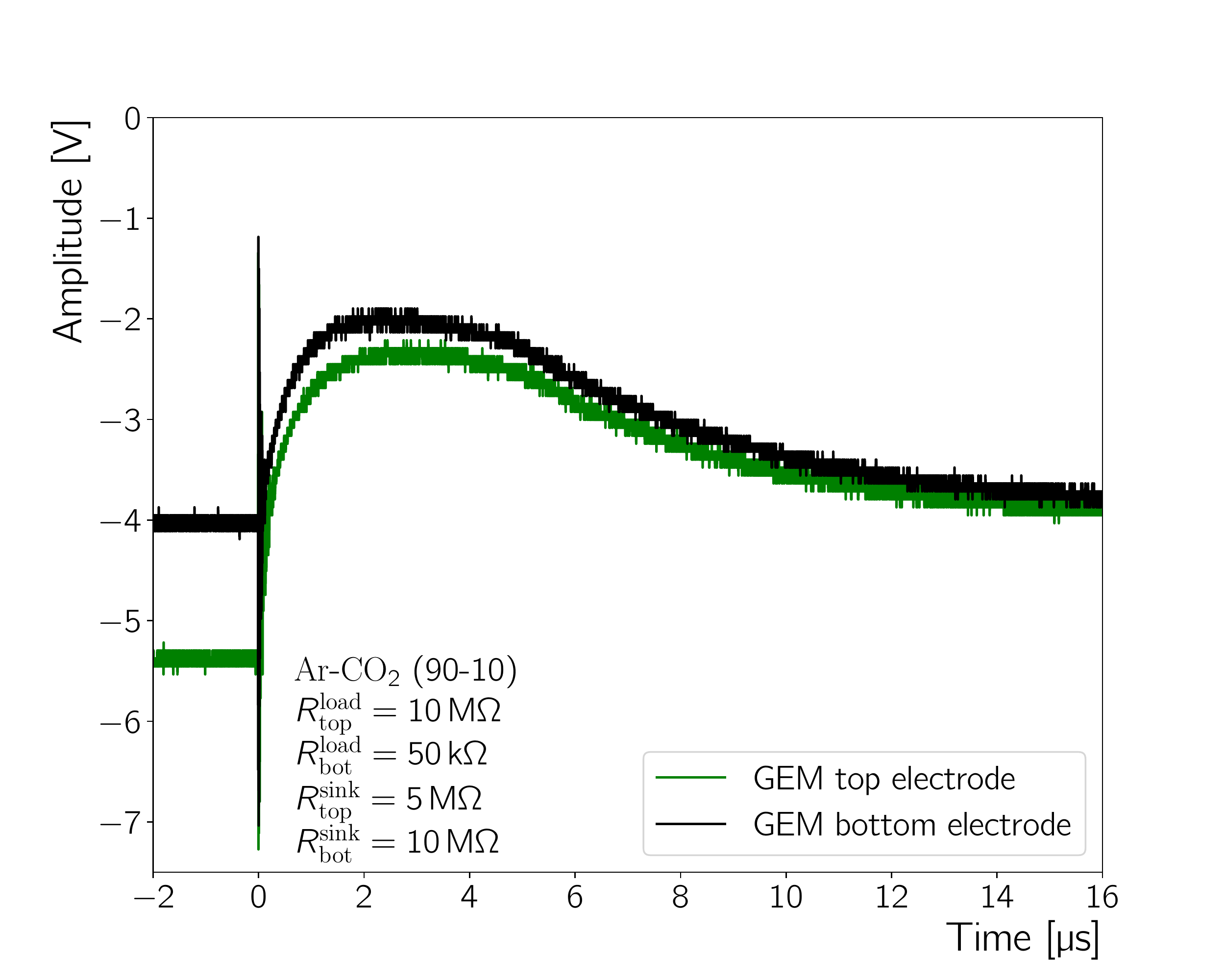}
\caption{\label{sec:results:fig:HVprobeSignals:SingleGEM:decouplingRes:noSec}(Colour online) High-voltage probe signals of a primary discharge. The waveforms have been recorded at an induction field of \SI{6.75}{\kilo\volt\per\centi\meter}.}
\end{figure}
\paragraph{The effect of decoupling resistors on the GEM potentials}$\ $\\
We examine HV probe signals in order to understand how the mitigation process via decoupling resistors works. These signals show that introducing a decoupling resistor affects the GEM potentials in the first $\sim\,\SI{10}{\micro\second}$ after the discharge. \Figref{sec:results:fig:HVprobeSignals:SingleGEM:decouplingRes:noSec} shows the time evolution of the potentials at the GEM top and bottom electrode after a discharge when $R_{\textrm{bot}}^{\textrm{load}}=\SI{50}{\kilo\ohm}$. As in the case of $R_{\textrm{bot}}^{\textrm{load}}=0$ (\secrefbra{sec:results:subsec:HVProbeSignals}, \figrefbra{sec:results:fig:HVprobeSignals:SingleGEM:noSecAndSec}) the GEM top potential drops at the time of the discharge and $\Delta V_{\textrm{GEM}}$ is thus significantly reduced. However, in contrast to the case without decoupling resistor (\figrefbra{sec:results:fig:HVprobeSignals:SingleGEM:noSecAndSec}), both GEM potentials continue to drop if a decoupling resistor is present (\figrefbra{sec:results:fig:HVprobeSignals:SingleGEM:decouplingRes:noSec}). This drop is not as abrupt as the drop of the GEM top potential at the time of a discharge ($t\sim\,0$). After about \SI{5}{\micro\second} the potentials increase again until they reach similar values as observed for the set-up without decoupling resistor. This additional voltage drop effectively reduces $E_{\textrm{ind}}$ below the threshold needed to ignite a secondary discharge. This happens at a time scale that seems important for the build up of a secondary discharge, judging from the times measured between primary and secondary discharge (\figrefbra{sec:results:fig:norm_vs_revers_times}). To balance this voltage drop at $R_{\textrm{bot}}^{\textrm{load}}$ and in order to observe secondary discharges, higher potential difference across the gap needs to be supplied.\\ \indent
Furthermore, the observation of a voltage drop across $R_{\textrm{bot}}^{\textrm{load}}$ indicates the presence of a current through the induction gap that decreases and vanishes with increasing time. Together with the current observed on the readout anode (\secrefbra{sec:results:subsec:currentsBefore2NDary}, \figrefbra{sec:results:fig:readoutPlaneSignalsHighAndLowResistanceToGND:noSec}) this HV probe measurement shows that there is indeed a current through the induction gap, which develops after the discharge and persists for some \si{\micro\second}.
\begin{figure*}
\centering
\subfloat[]{\includegraphics[width=0.49\textwidth, bb = 0 0 720 538, clip = true]{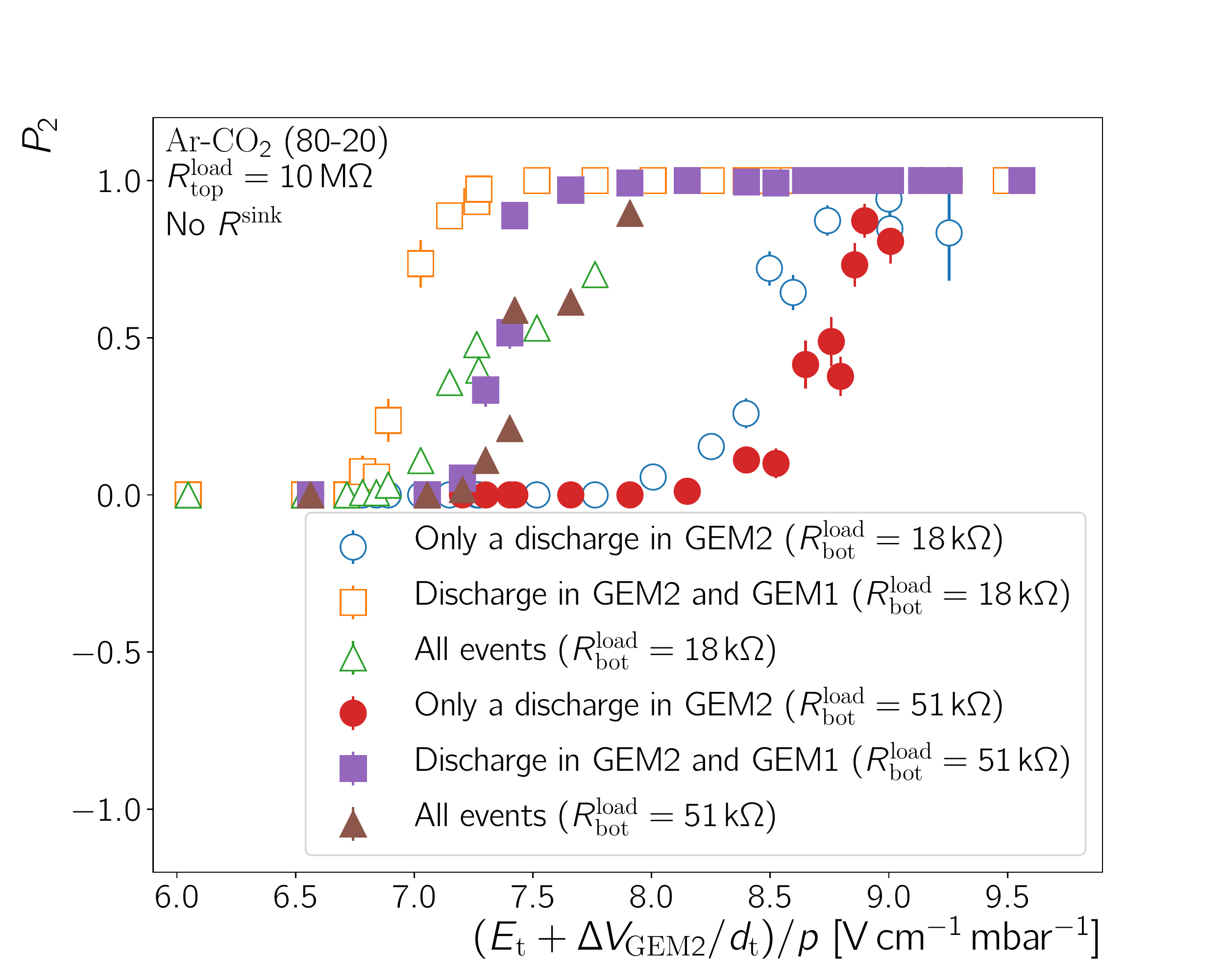}
\label{sec:results:fig:P2rDec18and51kOhm}}
\subfloat[]{\includegraphics[width=0.49\textwidth, bb = 0 0 720 538, clip = true]{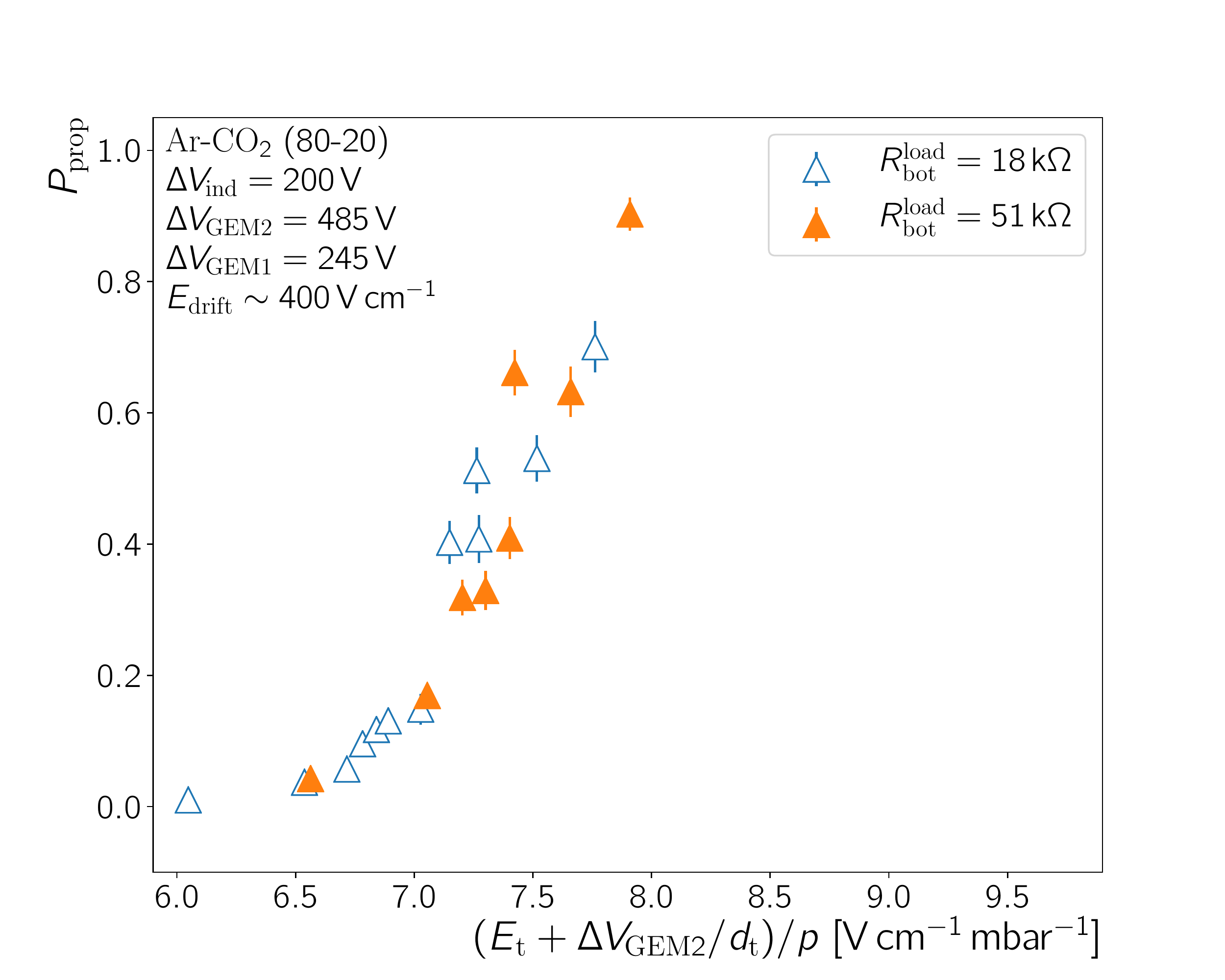}
\label{sec:results:fig:PProprDec18and51kOhm}}
\caption{\label{sec:results:fig:P2PProprDec18and51kOhm}(Colour online) Secondary discharge probability ($P_2$) and discharge propagation probability ($P_{\textrm{prop}}$) in the transfer gap between GEM2 and GEM1. Primary discharges are always induced in GEM2. \protect\subref{sec:results:fig:P2rDec18and51kOhm} $P_2$ calculated for all measured discharges and for two subgroups of all the events: (I) Events with only a discharge in GEM2 and (II) Events with a discharge in GEM2 and a propagated discharge in GEM1. \protect\subref{sec:results:fig:PProprDec18and51kOhm} Probability to propagate a discharge from GEM2 to GEM1, calculated from the data in \protect\subref{sec:results:fig:P2rDec18and51kOhm}. The settings indicated in each of the figures are valid for both of them.}
\end{figure*}

\subsection{Onset of secondary discharges in the transfer gap}
\label{sec:results:subsec:onsetInTransferGap}
Secondary discharges in the transfer gap feature a similar onset curve as secondary discharges in the induction gap, see \figref{sec:results:fig:rBotLoadVsET}. The transfer field $E_{\textrm{t}}$ is increased due to the drop of the GEM2 top electrode potential (\figrefbra{sec:results:fig:HVprobeSignals:readoutAndGEMPotentialsSecDc:etgem2tgem1b}) after the primary discharge. We therefore display measurements of secondary discharge in the transfer gap using the quantity $\overline{E}_{\textrm{t}} = E_{\textrm{t}} + \Delta V_{\textrm{GEM2}}/d_{\textrm{t}}$, where $\Delta V_{\textrm{GEM2}}$ and $d_{\textrm{t}}$ are the voltage difference between the GEM2 electrodes and the width of the transfer gap. $\overline{E}_{\textrm{t}}$ does not exactly correspond to the actual transfer field present just before a secondary discharge. However, this quantity allows to relate all the relevant set voltages to the measured parameters as \textit{e.g.} $P_2$. Furthermore $\overline{E}_{\textrm{t}}$ is closer to the actual field before the secondary discharge than $E_{\textrm{t}}$.\\ \indent
Measurements in \figref{sec:results:fig:rBotLoadVsET} are done with primary discharges in GEM2 and with a discharge propagation probability ($P_{\textrm{prop}}$) of 1. This means that all primary discharges in GEM2 are followed by a discharge in GEM1. Discharges with $P_{\textrm{Prop}}<1$ are studied for the remainder of this section in order to determine whether the behaviour of secondary discharges in the transfer gap is different when only GEM2 or GEM2 and GEM1 discharge. To this end the potentials at the GEM2 and GEM1 top electrode are recorded with the HV probes in order to determine the occurrence of secondary and GEM1 discharges.\footnote{In some other works the term \textit{propagated discharge} (\textit{e.g.} \cite{Bachmann}) is as well used for the phenomenon that we describe as secondary discharge. In this work we clearly describe different phenomena with the two words, \textit{i.e.} the discharge of the gap between GEMs (secondary discharge) and the propagation of a discharge from one GEM to another (discharge propagation).}\\ \indent
When a  discharge in GEM1 occurs, secondary discharges are observed at significantly lower $E^{\textrm{on}}_{\textrm{t}}$ as compared to the case where only GEM2 discharges (\figrefbra{sec:results:fig:P2rDec18and51kOhm}). The onset fields for the two cases differ by about \SI{1.5}{\volt\per\centi\meter\per\milli\bar}. The mitigating effect of the decoupling resistor is well visible for events with a GEM1 discharge, however, for events with no discharge the mitigation is weaker as the comparison between the $R_{\textrm{bot}}^{\textrm{load}}=\SI{18}{\kilo\ohm}$ and \SI{51}{\kilo\ohm} measurement shows.\\ \indent
We calculate the propagation probability $P_{\textrm{prop}}=N_{\textrm{GEM1}}/N_{1}$ from the data in \figref{sec:results:fig:P2rDec18and51kOhm} using the number of events with a propagated discharge to GEM1 ($N_{\textrm{GEM1}}$) and the number of primary discharges ($N_1$). \Figref{sec:results:fig:PProprDec18and51kOhm} shows that $P_{\textrm{prop}}$ increases with increasing transfer field and that there is no dependence on $R_{\textrm{bot}}^{\textrm{load}}$. Note that all other voltage differences have been kept constant as other works identified a dependence of $P_{\textrm{prop}}$ on the $\Delta V_{\textrm{GEM}}$ of the GEM the discharge is propagated to \cite{Bachmann,Deisting:2018gjp,baDatz2017}.\\ \indent
It is thus important not only to consider $P_2$ for two cases with and without GEM1 discharge, but the product $P_{\textrm{Prop}}\times P_2$. \textit{E.g.} in \arcois{80-20} (\figrefbra{sec:results:fig:P2rDec18and51kOhm}, \textit{All events $R_{\textrm{bot}}^{\textrm{load}}=\SI{18}{\kilo\ohm}$} and \textit{\SI{51}{\kilo\ohm}}) the particular $\overline{E}^{\textrm{on}}_{\textrm{t}}$ for events with and without GEM1 discharge and the dependence of $P_{\textrm{prop}}$ on $\overline{E}_{\textrm{t}}$ lead to a situation where the $P_2$ onset curves for all secondary discharges are driven by the discharge propagation (Figs.~\ref{sec:results:fig:P2rDec18and51kOhm} and \ref{sec:results:fig:PProprDec18and51kOhm}). In conclusion, minimising the discharge propagation between two GEMs allows to reach higher transfer fields without the occurrence of secondary discharges.

%% file: formationOfSecondaryDischarges.tex
\section{Formation of secondary discharges}
\label{sec:formationSecDCs}
\begin{figure*}
\centering
\subfloat[]{\label{sec:formationSecDCs:fig:e-energyDist}\includegraphics[width=0.49\textwidth, bb = 0 0 720 538, clip = true]{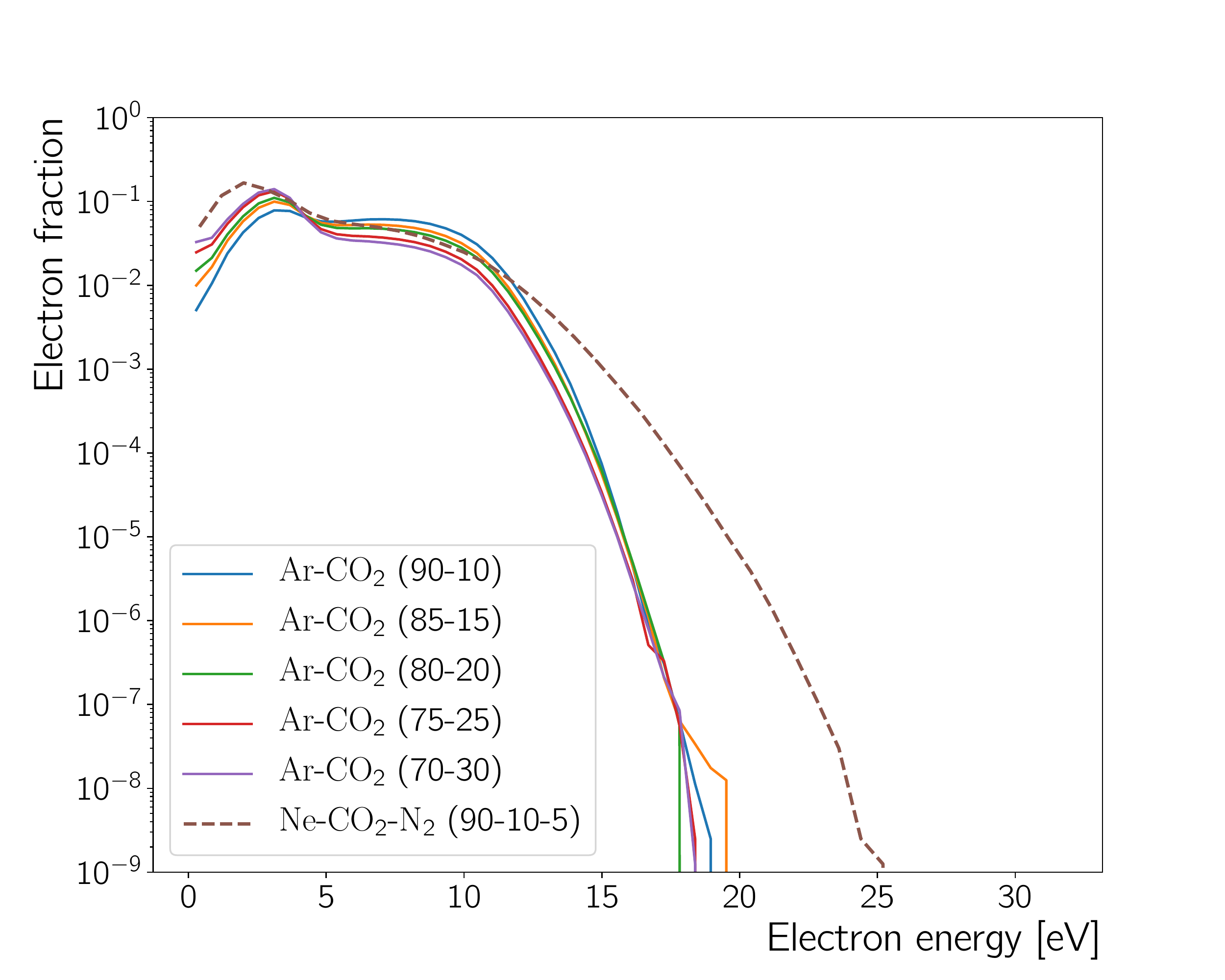}}
\subfloat[]{\label{sec:formationSecDCs:fig:waveform_and_fit}\includegraphics[width=0.49\textwidth, bb = 0 0 720 538, clip = true]{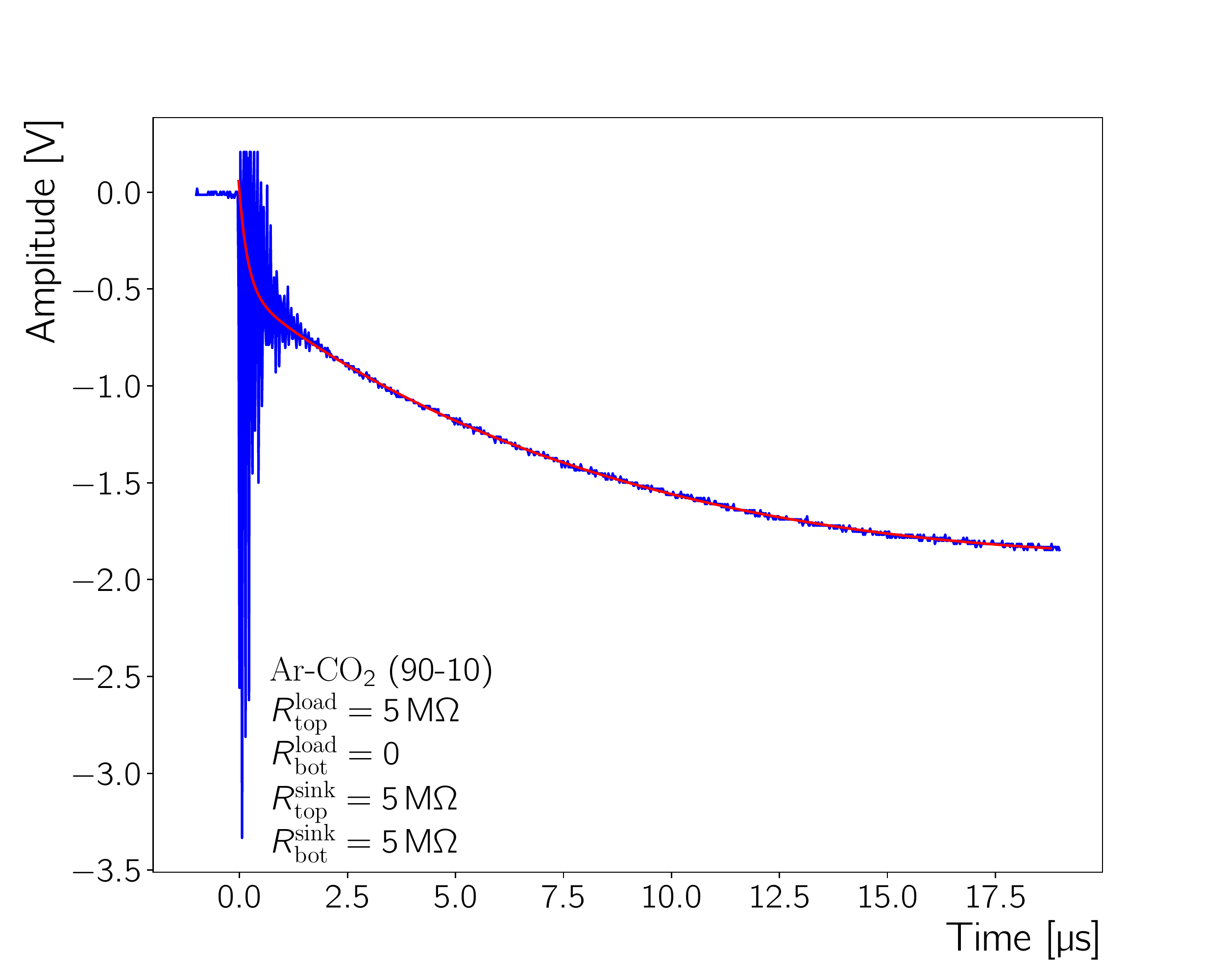}}
\caption{\label{sec:formationSecDCs:fig:e-energyDist_and_waveform_and_fit}(Colour online) \protect\subref{sec:formationSecDCs:fig:e-energyDist} Normalised electron energy distribution at the onset field as simulated with Magboltz. The $E_{\textrm{ind}}^{\textrm{on}}$ values form Table \ref{sec:results:tab:onset_for_different_mixtures}, pressure of \SI{953.3}{\milli\bar} and a temperature of $\SI{23}{^{\circ}C}$ have been used. \protect\subref{sec:formationSecDCs:fig:waveform_and_fit} Measurement of the anode potential (blue) during a primary discharge in the GEM and a fit to the measurement (red).}
\end{figure*}
\begin{figure*}
\centering
\subfloat[]{\includegraphics[width=0.49\textwidth, bb = 0 0 720 538, clip = true]{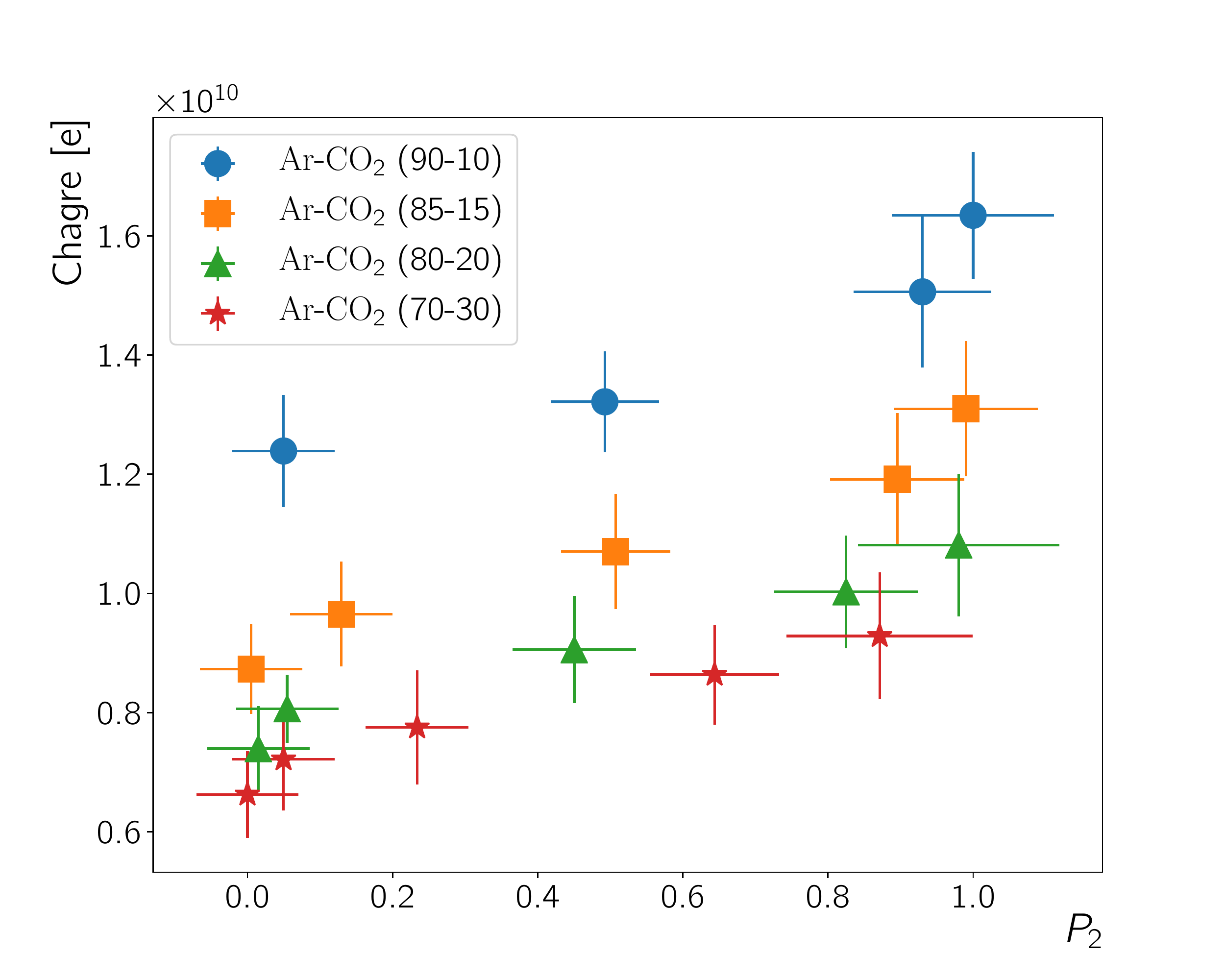}\label{sec:formationSecDCs:fig:charge_vs_sec_prob}}
\subfloat[]{\includegraphics[width=0.49\textwidth, bb = 0 0 720 538, clip = true]{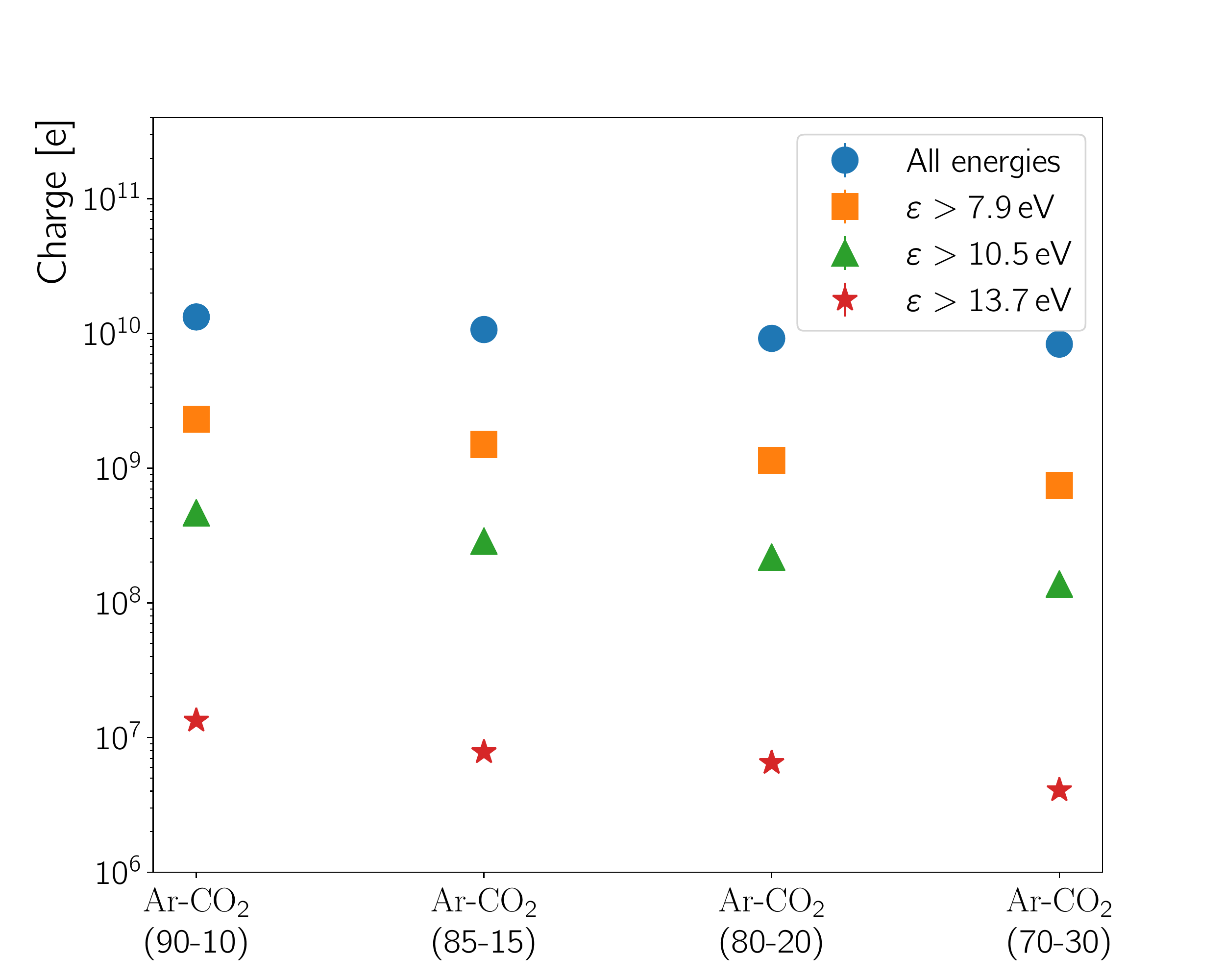}\label{sec:formationSecDCs:fig:fractions_vs_co2}}
\caption{\label{sec:formationSecDCs:fig:fractions_vs_co2-and-charge_vs_sec_prob}(Colour online) \protect\subref{sec:formationSecDCs:fig:charge_vs_sec_prob} Measurement of the effective charge produced by the primary discharge in the GEM. \protect\subref{sec:formationSecDCs:fig:fractions_vs_co2} Number of electrons produced during the primary discharge for different electron energies ($\varepsilon$).}
\end{figure*}
Secondary discharges are characterised by the following features: they happen a few tens of \si{\micro\second} after a primary discharge (\textit{e.g.} \figrefbra{sec:results:fig:norm_vs_revers_times}); during this intermediate time, a current flowing through the concerned gap is observed (\textit{cf.} \secrefbra{sec:results:subsec:currentsBefore2NDary}); and the Townsend coefficient at the onset fields for secondaries is zero.\\ \indent
However, the tails of the electron energy distributions upon a primary discharge might result in further ionization and excitation of the gas. We performed Magboltz simulations in order to obtain the electron distributions at $E^{\textrm{on}}_{\textrm{ind}}$. The onset field values are extracted from our measurements, as discussed in \secref{sec:results:subsubsec:differentMixtures}, for various concentrations of $\textrm{CO}_{2}$ in $\textrm{Ar}$ and also for \baseline{}.\\ \indent
The output of these simulations, depicted in \figref{sec:formationSecDCs:fig:e-energyDist}, shows that indeed there is a small fraction of electrons with energies high enough to excite and even ionise the gas, considering that the ionisation potentials of $\textrm{CO}_{2}$ and $\textrm{Ar}$ are \SI{13.7}{\electronvolt} and  \SI{11.8}{\electronvolt}, respectively \cite{Biagi1018382}.\\ \indent
In order to define an absolute number of electrons produced during a primary discharge across the GEM, an additional set of measurements is carried out for different $\textrm{CO}_{2}$ concentrations in \arco{} mixtures. The experimental set-up is similar to the one described in \secref{sec:setup}, with the only difference that the anode signal is read out with a standard oscilloscope probe. The readout anode is grounded through a parallel RC circuit (R=\SI{100}{\kilo\ohm}, C=\SI{100}{\nano\farad}), analogue to the measurements described in \cite{Antonija}.\\ \indent
A statistically significant number of waveforms is recorded for \arco{} gas mixtures with \SI{10}{\%}, \SI{15}{\%}, \SI{20}{\%} and \SI{30}{\%} $\textrm{CO}_{2}$ concentration at several values of the induction field within the range of the onset curve for the given gas mixture. A typical waveform recorded during primary discharges is shown in \figref{sec:formationSecDCs:fig:waveform_and_fit}. An effective charge that induces current in the readout circuit can be defined as an integral of this current over time $\int_{0}^{t_{\textrm{eff}}}I(t)\;\textrm{d}t$, where $t_{\textrm{eff}}$ is the time required for the initial current to decrease down to $1/e$ of its original value. This criterion is used since the waveform may or may not include a secondary discharge, so not the whole waveform can be fitted for this purpose. The current $I(t)$ is the derivative of the voltage $V(t)$ across the readout capacitor with respect to time, multiplied by the value of the capacitance C. In order do define $V(t)$ all the waveforms are fitted with the exponential function $\sum_{i=1}^{n}k_i\;(1-e^{-\frac{t}{\tau_i}})$ \cite{Antonija} where $\textrm{n}=5$ and the coefficients $k_{\textrm{i}}$ and $\tau_{\textrm{i}}$ are extracted from a non-linear least-squares method.\\ \indent
\begin{figure}
\centering
\includegraphics[width=0.49\textwidth, trim = 0 0 0 38, bb = 0 0 720 538, clip = true]{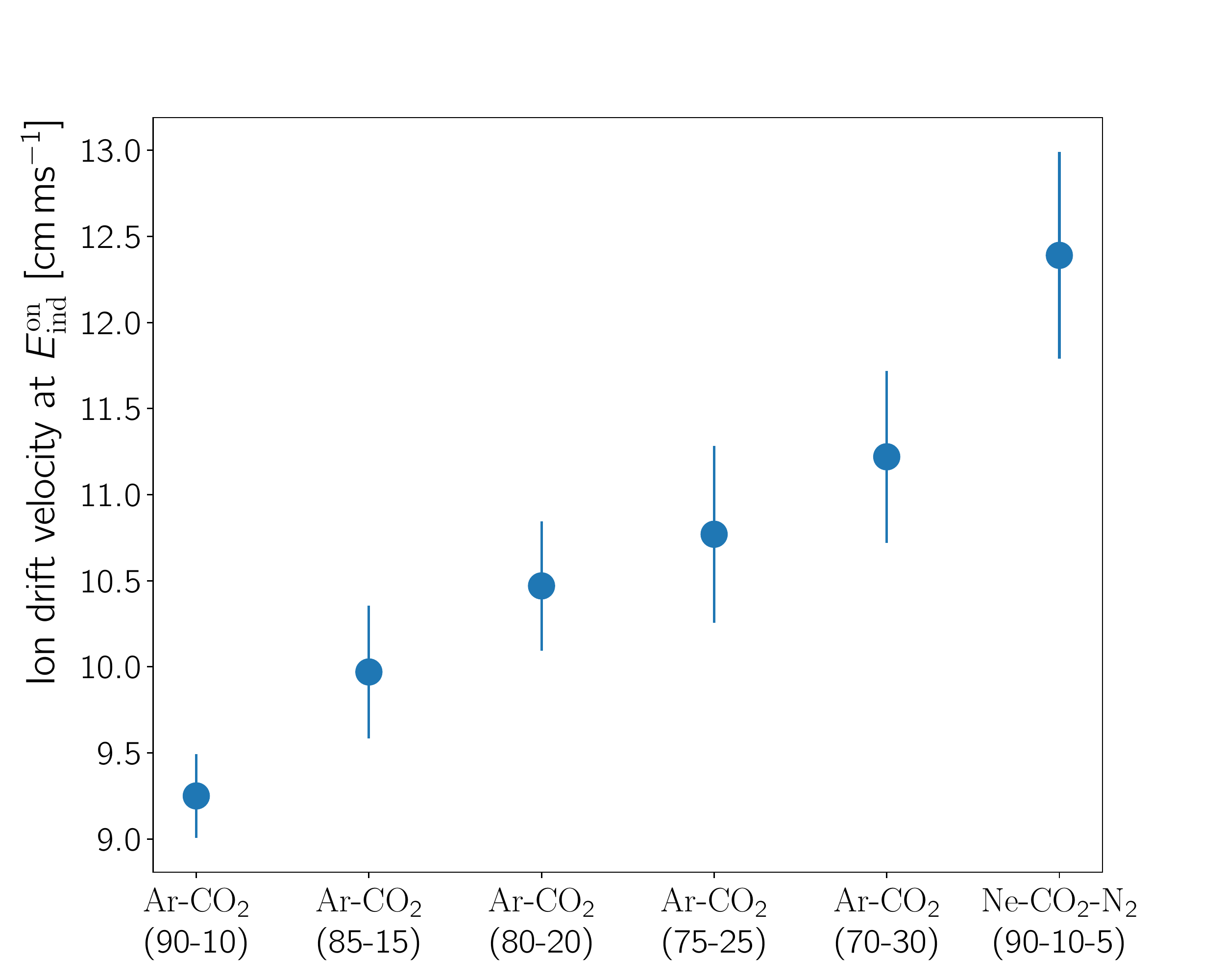}
\caption{\label{sec:formationSecDCs:fig:ionVelocities}Calculated ion drift velocity as function of the gas mixture at the onset field. To calculate these values the mobilities in \cite{DEISTING20181} have been multiplied with the $E_{\textrm{ind}}^{\textrm{on}}$ extracted from \figref{sec:results:fig:onset_for_different_mixtures:daniel}, \textit{cf.} Table \ref{sec:results:tab:onset_for_different_mixtures}.} 
\end{figure}
An effective charge is calculated for several induction fields for a given \arco{} gas mixture (see \figrefbra{sec:formationSecDCs:fig:charge_vs_sec_prob}). In order to relate the obtained absolute number of electrons with the fractions of the electrons available for gas excitation/ionisation provided by Magboltz, the exact values of the effective charge at $E_{\textrm{ind}}^{\textrm{on}}$ are defined by means of linear interpolation of the experimental data points in the figure. The number of electrons with an energy higher than the excitation and ionization potentials of $\textrm{Ar}$ atoms and $\textrm{CO}_{2}$ molecules is thus defined by the product of the effective charge and the corresponding normalised fractions.\\ \indent
The total number of electrons and the number available for excitation/ionization of $\textrm{CO}_{2}$ and $\textrm{Ar}$ as a function of the $\textrm{CO}_{2}$ concentration is shown in \figref{sec:formationSecDCs:fig:fractions_vs_co2}. The first and the second excited states of $\textrm{CO}_{2}$ (\SI{7.9}{\electronvolt} and \SI{10.5}{\electronvolt}), the first and the second excited states of $\textrm{Ar}$ (\SI{11.3}{\electronvolt}, \SI{11.6}{\electronvolt}) and the ionization potentials of $\textrm{CO}_{2}$ and $\textrm{Ar}$ (\SI{13.7}{\electronvolt}, \SI{11.8}{\electronvolt}) are taken into account \cite{Biagi1018382}. The number of available electrons decreases with $\textrm{CO}_{2}$ concentration, but stays in the same order of magnitude.\\ \indent
This study reveals that in fact the amount of charge needed to trigger a secondary discharge seems to be well defined for a given gas mixture, but shows a slight dependence on the exact composition. On the other hand the positively charged ions produced after the primary discharge should be also taken into account. It has been already demonstrated in \secref{sec:results:subsubsec:influenceOfTheDriftField} that the ions that are reaching the GEM top layer are not responsible for the appearance of secondary discharges. However, the ions hitting the GEM bottom side could play a role by causing extraction of secondary electrons from the GEM surface or from the electrode which plays the role of a cathode. From the known values of the ion mobilities \cite{DEISTING20181}, it is possible to calculate the drift velocity of the ions at the onset fields, which is a proxy for the ions' energy \cite{masontransport}. \Figref{sec:formationSecDCs:fig:ionVelocities} shows that indeed with increasing $\textrm{CO}_2$ in $\textrm{Ar}$, the ion velocity, and thus their energy, increases. Therefore, with higher $\textrm{CO}_2$ concentrations, less electrons, producing higher energy ions, lead to secondary discharges. Note that in the case of reverse fields, where ions must travel through the gap in order to reach the cathode, the sheer amount of drifting charges may create a strong space-charge accelerating field.\\ \indent
These results point to heating of the cathode, upon a primary discharge, as the possible mechanism which ultimately ends with a secondary discharge.
If enough charge is produced in the primary discharge, further ionisation and excitation produce thermionic emission of electrons from the cathode, which is in turn heated by ion bombardement. This process is self-sustained until eventually massive electron emission results in the breakdown of the gap.

%% file: summary.tex
\section{Conclusion}
\label{sec:summary}
A GEM set-up with either one or two $10\times\SI{10}{\centi\meter\squared}$ GEMs has been used to study secondary discharges in the induction gap between GEM2 and the readout anode and -- in the double GEM case -- in the transfer gap between GEM2 and GEM1. In this paper we extend the existing knowledge on secondary discharges and their phenomenology \cite{Peskov,Bachmann} and provide new insights into the subject.\\ \indent
Secondary discharges occur exclusively after a primary discharge in a GEM and correspond to a breakdown of the respective gap. Their occurrence probability as a function of the induction or transfer field rises over a narrow field range from 0 to 1, where their onset field is lower than the electric field needed for Townsend amplification. No increase of the actual electric field in the gap is observed before a secondary discharge, therefore such an increase can not explain its occurrence. As expected by ionisation physics, the onset field for secondary discharges increases with increasing fraction of the quencher.\\ \indent
A current is observed at the readout anode after the primary discharge in the GEM, which decays over $\sim\SI{10}{\micro\second}$. Furthermore, a voltage drop at the resistor in the HV supply line to the GEM bottom electrode ($R_{\textrm{bot}}^{\textrm{load}}$) is measured, indicating a current, too. We conclude from these two results that there is a current through the gas in the induction gap after the primary and prior to a possible secondary discharge. With Gas Discharge Tubes (GDTs) we decouple the actual discharge from the counting gas in our detector. The GDT studies show that secondary discharges are only observed if a primary discharge happens in the GEM and thus in the counting gas. The same holds for the current after the primary discharge. We therefore conclude that secondary discharges are caused by an effect in the gas and not by RCL effects in the HV supply circuit or the response of the supply circuit to the primary discharges.\\ \indent
Secondary discharges in the induction and the transfer gap are found to evolve along and against the direction of the electric field. The modulus of the electric field is the driving factor for the onset of secondary discharges in the induction gap. The time between primary and secondary discharge ($t_2$) for both field directions is similar and not compatible with the electron velocity. The time $t_2$ decreases with field intensity faster than the ion drift velocity, thus ruling out secondary discharge mechanisms relying only on charge carriers crossing the gap.\\ \indent
Furthermore, the ion extraction from the GEM holes after a discharge seems irrelevant for the creation of secondary discharges in the induction gap. Varying the drift field in the gap on the GEM's side opposite to the readout anode from $-\SI{290}{\volt\per\centi\meter}$ to \SI{442}{\volt\per\centi\meter} neither affects $P_2$ nor the time between primary and secondary discharge, although such a variation affects the ion extraction efficiency.\\ \indent
Introducing a decoupling resistor into the HV supply path of the bottom GEM electrode allows to mitigate the occurrence of secondary discharges. The electric field at which secondary discharges appear increases with increasing $R^{\textrm{load}}_{\textrm{bot}}$. We observe an onset field increase of about \SI{0.1}{\volt\per\centi\meter\per\milli\bar} for an increase in resistance of \SI{10}{\kilo\ohm}. The exact value differs for individual GEMs and gas mixtures. Thus, $R^{\textrm{load}}_{\textrm{bot}}$ provides a feasible way of mitigating the occurrence of secondary discharges during operation of a detector employing GEMs, since the current through this resistor, after a discharge, leads to an effective field reduction in the affected gap.\\ \indent
Propagation of primary discharges from one GEM to anotheraffects secondary discharges in the transfer gap between these two GEMs. The probability to propagate a discharge from GEM2 to GEM1 rises with increasing transfer field. In case a primary discharge is propagated from GEM2 to GEM1, secondary discharges are observed at an onset field lower by \SI{1.5}{\volt\per\centi\meter\per\milli\bar} than in the case of no propagation. Therefore mitigation of discharge propagation also mitigates the occurrence of secondary discharges.\\ \indent
Based on the measurements and simulations presented in this work, we propose a mechanism responsible for the occurrence of secondary discharges: upon the primary discharge, a sufficient amount of electrons are produced that can ionise and excite the gas in the concerned gap. The primary discharge produces also a certain heat load on the GEM electrodes, which in turn facilitates the thermionic emission of further electrons upon ion bombardment and infra-red radiation. The total charge produced after the primary discharge seems to lead the fate of this self-sustained currents into a secondary discharge. This characteristic charge is found to be on the order of $10^{10}$ electrons and to slightly decreases with increasing quencher, probably due to the increasing strength of ion bombardment on the cathode surface.